\documentclass[a4paper,fleqn,usenatbib]{mnras}

\usepackage{newtxtext,newtxmath}
\usepackage{gensymb}

\usepackage[T1]{fontenc}
\usepackage{ae,aecompl}
\usepackage{soul}	

\usepackage{graphicx}	
\usepackage{amsmath}	
\usepackage{amssymb}	
\usepackage{color}

\title[Secondary's rotation in synchronous binary]{On the Secondary's Rotation in Synchronous Binary Asteroid}

\author[H.S. Wang and X.Y. Hou]{
H.S. Wang,$^{1,2,3}$
X.Y. Hou,$^{1,2,3}$\thanks{E-mail: houxiyun@nju.edu.cn}
\\
$^{1}$School of Astronomy and Space Science, Nanjing University, Nanjing 210093, China\\
$^{2}$Institute of Space Environment and Astronautics, Nanjing University, Nanjing 210093, China\\
$^{3}$Key Laboratory of Modern Astronomy and Astrophysics, Nanjing University, Ministry of Education, Nanjing 210093, China
}

\date{Accepted XXX. Received YYY; in original form ZZZ}

\pubyear{2015}

\begin{document}
\label{firstpage}
\pagerange{\pageref{firstpage}--\pageref{lastpage}}
\maketitle

\begin{abstract}
	This paper studies the secondary's rotation in a synchronous binary asteroid system in which the secondary enters the 1:1 spin-orbit resonance. The model used is the planar full two-body problem composed of a spherical primary plus a tri-axial ellipsoid secondary. Compared with classical spin-orbit work, there are two differences: (1) Influence on the mutual orbit from the secondary's rotation is considered; (2) Instead of the Hamiltonian approach, the approach of periodic orbits is adopted. Our studies find: (1) Genealogy of the two families of periodic orbits is same as that of the families around triangular libration points in the restricted three-body problem. That is, the long-period family terminates onto a short-period orbit travelling $N$ times; (2) In the limiting case where the secondary's mass is negligible, our results can be reduced to the classical spin-orbit theory, by equating the long-period orbit with the free libration, and by equating the short-period orbit with the forced libration caused by orbit eccentricity. However, the two models show obvious differences when the secondary's mass is non-negligible. (3) By studying the stability of periodic orbits, for a specific binary asteroid system, we are able to obtain the maximum libration amplitude of the secondary (which is usually less than $90 \degree$), and the maximum mutual orbit eccentricity which does not break the secondary's synchronous state. We also find the anti-correlation between the secondary's libration amplitude and the orbit eccentricity. The (65803) Didymos system is taken as an example to show the results.
\end{abstract}




\section{Introduction}

The synchronous state is widespread in the solar system, from the Sun-planet systems to the planet-satellite systems \citep{Murray1999}. Due to tidal dissipation \citep{Goldreich1966}, given enough time, the eventual fate of two bodies circling each other is the synchronous state. If the satellite is trapped in such a state, its rotation period is same as the mutual orbit period, so it has its one-side always facing the other body, such as the Moon in our Earth-Moon system. In literature, we also call the synchronous state as being trapped in the 1:1 spin-orbit resonance. Except the 1:1 one, there are other spin-orbit resonances in which one body’s rotation period is commensurate with its orbital period. Nevertheless, till now Mercury is the only natural celestial body in the solar system which is confirmed to be trapped in a spin-orbit resonance other than the 1:1 one.

The inherent dynamics of spin-orbit resonances is spin-orbit coupling, a mechanism by which the satellite’s rotation and its orbital motion influence each other. There are tremendous work on the spin-orbit coupling problem, from specific resonances to chaotic rotations. To list only a few, please see \citep{Wisdom1984,Celletti1990a,Celletti1990b,Noyelles2014,Quillen2017}. Most of these studies assume an invariant mutual orbit, i.e., neglect the influence from the satellite's rotational motion on the orbital motion. This assumption is valid, in the sense that the satellite is usually much smaller compared with the planet, and its size is much smaller compared with the mutual orbit distance. However, this assumption is challenged when studying spin-orbit coupling in binary asteroid systems. The two asteroids usually are highly non-spherical, close to each other, and may have masses comparable to each other. Usually there are strong spin-orbit coupling in binary asteroid systems \citep{Maciejewski1995,Scheeres2004,Scheeres2006Dynamical,Fahnestock2006,Chappaz2015,Ferrari2016,Hou2017,Yu2017,Hirabayashi2019}. The assumption of an invariant mutual orbit may cause qualitative difference from the real physics when studying spin-orbit coupling in these systems \citep{Hou2017,Hou2017A}.

Till now, only three states of binary asteroid system are confirmed \citep{Margot2015,Walsh2015,Pravec2016}: (1) the doubly synchronous state in which both the primary and the secondary are trapped in 1:1 spin-orbit resonance; (2) the synchronous state in which only the secondary is trapped in 1:1 spin-orbit resonance; (3) the asynchronous state in which both asteroids are not in 1:1 spin-orbit resonance. Except a few candidates of the doubly synchronous state, the majority of the discovered binaries are either in the synchronous state or the asynchronous state. Since our focus is spin-orbit resonances, this study is devoted to the secondary’s rotation in a synchronous binary asteroid system. By simultaneously considering the orbital and rotational motions: (1) We try to answer the question that to what extent the mutual orbit eccentricity and the libration amplitude can be for a specific synchronous binary system. (2) In the case that the secondary’s mass is much smaller than the primary, we find that the results of the classical spin-orbit theory can be recovered from our model. However, when the secondary’s mass is non-negligible, the difference between the two models is obvious. (3) We find the anti-correlation between the orbit eccentricity and the libration amplitude. For a specific binary asteroid system, the larger the orbit eccentricity is, the smaller the maximum libration amplitude is. These findings can help us understand the phenomenon that synchronous binary asteroid systems generally have small orbit eccentricities \citep{Pravec2016}. One remark is that in this study we only consider the mutual gravity between the two asteroids, neglecting the long-term tidal and thermal effects.

Different from previous work which usually reduces the Hamiltonian to the form of a perturbed simple pendulum, we carry out our work in the approach of periodic orbits. This approach is often used to study orbital resonances in the solar system \citep{Broucke1969,Hadjidemetriou1975}, but seldom used in the study of spin-orbit resonances. We find that this approach is ideal for the purpose of this work. That is to find the maximum libration amplitude or maximum orbit eccentricity for the stable 1:1 spin-orbit resonant configuration. The model used in this study is simple, composed of a sphere primary and an ellipsoidal secondary. By choosing a sphere primary we can exclude the primary’s rotation from our study and we only focus on the secondary’s rotation. This model has been used by some previous researchers on general dynamics of binary asteroid systems \citep{Bellerose2008,Gabern2006,Mcmahon2010}.

The paper is organized as follows. Section 2 introduces the dynamical model used in this study and equations of motion (EOMs) in the secondary’s body-fixed frame. For the exact 1:1 spin-orbit resonance, the primary in the secondary’s body-fixed frame is stationary at equilibrium points (EPs). Section 3 studies the stability of these EPs. Focusing on the stable EPs, section 4 studies the two basic families of periodic orbits. We find that genealogy of the two families is same as that of the long and the short-period family around triangular libration points of the restricted three-body problem \citep{Henrard2002,Hou2009bridges}, and also same as that of the two planar families around the equilibrium points in the body-fixed frame of a uniformly rotating asteroid \citep{Feng2017,Jiang2019}. In section 5, we relate our model with the classical spin-orbit resonance model, and relate our mathematical results with physics of the binary asteroid systems. In this work, we mainly focus on the binary (65803) Didymos system, but some general analysis is also made.
 
\section{Model Description}
\label{sec:Model}

\begin{figure}
		\includegraphics[width=\columnwidth]{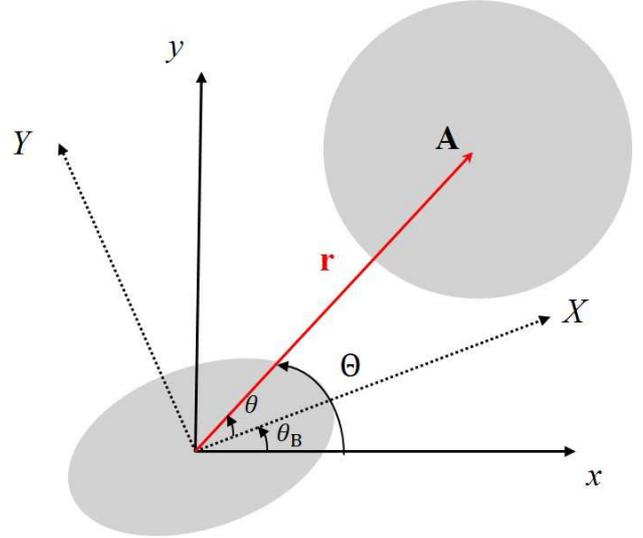}
		\caption{Relative geometry of the planar sphere+ellipsoid model used in this work.}
		\label{fig:Model}
	\end{figure}

	The dynamical model used in this paper is described in this section. Since we focus on the spin-orbit resonance of the secondary, we use the simple model composed of a sphere primary and a tri-axial ellipsoid secondary. Further, we assume that the secondary rotates along its shortest axis and its equator coincides with its orbital plane. As a result, the problem is planar. Fig.\ref{fig:Model} shows the geometry of the system.
	The coordinate system $O-XY$ is the body-fixed frame of the secondary $B$. The inertial frame $O-xy$ also has its origin at the secondary. $\theta_{B}$ is {$B$}'s rotation angle. $\theta$ is the phase angle of the primary's position vector in {$B$}'s body-fixed frame, and {$r$} is the distance between the secondary and the primary. The angles satisfy equations:
	\begin{equation}
		\Theta = \theta_{B}+\theta 
		\label{eq:angles_relation}
	\end{equation}
	The semi-axis and masses of secondary and primary are denoted as
	\begin{equation}
		a_{B},b_{B},c_{B},m_{B}; \quad r_{A},m_{A}; \notag
		\label{eq:axis_mass_icon}
	\end{equation}
	Non-spherical gravity coefficients $J_2^{B}$ and $J_{22}^{B}$ of the secondary are given by \citep{Balmino1994}
	\begin{equation}
		J_2^{B}=\frac{a_{B}^2+b_{B}^2-2c_{B}^2}{10\overline{a}_{B}^2} \qquad J_{22}^{B}=\frac{a_{B}^2-b_{B}^2}{20\overline{a}_{B}^2} \notag
	\end{equation}
	The reference radius $\overline{a}_{B}$ is defined as 
	\begin{equation}
		\overline{a}_{B} = (a_{B}b_{B}c_{B})^{1/3}\\
	\end{equation}
	Following units (mass [M], length [L], time [T] and energy [E]) are used in this work.
	\begin{equation}
		\begin{array}{l}
			{[M] = m_A+m_B \quad {[L]} = r_{A}+a_{B}} \\ 
			{[T] = \sqrt{\frac{{[L]}^3}{G[M]}} \qquad [E] = \frac{G[M]^2}{{[L]}}}\\
		\end{array}
	\label{con:unit}
	\end{equation}

	In the above units, the potential function $V$, system's energy $E$, and system's total angular momentum $K$ are
	\begin{equation}
		\begin{array}{l}
			{V = \frac{m}{r}+\frac{m}{2{r}^3}[A_1+A_2\cos(2\theta)]}\\
			{E = \frac{m}{2}[\dot{r}^2+{r}^2(\dot{\theta}+\dot{\theta}_{B})^2]+\frac{1}{2}I_z^{B}\dot{\theta}_{B}^2-V}\\
			{K = mr^2(\dot{\theta}_{B}+\dot{\theta})+I_z^{B}\dot{\theta}_{B}}
		\end{array}
	\label{con:lagrangian}
	\end{equation}
	in which
	\begin{equation}
		I_z^{B} = \frac{m_{B}(a_{B}^2+b_{B}^2)}{5[M]{[L]}^2}
	\end{equation}
	\begin{equation}
		\begin{array}{l}
			{\mu = \frac{m_{B}}{m_A+m_B} \quad\qquad m = \frac{m_Am_B}{[M]^2}}\\
			{A_1 = \alpha_{B}^2J_2^{B} \qquad\qquad A_2 = 6\alpha_{B}^2J_{22}^{B}}\\
			{\alpha_{B} = \overline{a}_{B}/{[L]}}
		\end{array}
	\label{con:A1A2}
	\end{equation}
	From Eq.\ref{con:lagrangian}, equations of motion are \citep{Hou2017A}:
	\begin{equation}
		\left\{
		\begin{array}{lr}
			\ddot{{r}} = {r}(\dot{\theta}+\dot{\theta}_{B})^2-\frac{1}{{r}^2}-\frac{3}{2{r}^4}[A_1+A_2\cos(2\theta)]\\
			\ddot{\theta} = -\frac{A_2}{{r}^5}\sin(2\theta)-2\frac{\dot{r}}{r}(\dot{\theta}+\dot{\theta}_{B})-\frac{mA_2}{I_z^{B}}\frac{\sin(2\theta)}{{r}^3}\\
			\ddot{\theta}_{B} = \frac{mA_2}{I_z^{B}{r}^3}\sin(2\theta)\\
		\end{array}
		\right.
	\label{con:EOM3}
	\end{equation}
	Note that we can use the conservation of the total momentum to reduce 1 DOF {(degree of freedom)} of Eq.\ref{con:EOM3}. According to the fourth equation of Eq.\ref{con:lagrangian}, by expressing $\dot{\theta}_B$ as a function of $K$, $\dot{\theta}$ and $r$, and by substituting the relation to the first two equations of Eq.\ref{con:EOM3}, we have
	\begin{equation}
	\left\{
		\begin{array}{lr}
		\ddot{{r}}={r}(\frac{I_z^{B}\dot{\theta}+K}{I_z^t})^2-\frac{1}{{r}^2}-\frac{3}{2{r}^4}[A_2\cos(2\theta)+A_1]\\
		\ddot{\theta}=-\frac{2\dot{{r}}(I_z^{B}\dot{\theta}+K)}{{r}I_z^t}-\frac{A_2\sin(2\theta)I_z^t}{I_z^{B}{r}^5}\\
				\end{array}
	\right.
	\label{con:EOM2}
	\end{equation}
	In which
	\begin{equation}
		I_{z}^{t}=I_{z}^{{B}}+m{r}^{2}
	\end{equation}
	Now, Eq.\ref{con:EOM2} is a 2-DOF dynamical system. Note that Eq.\ref{con:EOM2} is identical to Eqs.(18) and (19) in \citep{McMahon2013} where the non-spherical terms are expressed as the secondary's moments of inertia and different units are used. There is an energy integral for Eq.\ref{con:EOM2}, which is the energy integral given by the second equation of Eq.\ref{con:lagrangian}. We call the curve described by
	\begin{equation}
		E=\frac{K^2}{2(I_z^{B}+m{r}^2)}+\frac{m}{{r}}+\frac{3m}{2{r}^3}(A_1+A_2\cos(2\theta))
	\end{equation}
	as the zero velocity curve (Z.V.C.). Besides the Z.V.C., there are four EPs of Eq.\ref{con:EOM2} \citep{Wang2018}, which we will address in the following subsection.

\section{Exact 1:1 spin-orbit Resonance---Equilibrium Points}
\label{sec:SOR}

	\subsection{Position of the EPs}
\begin{figure}
			\centering
			\includegraphics[width=\columnwidth]{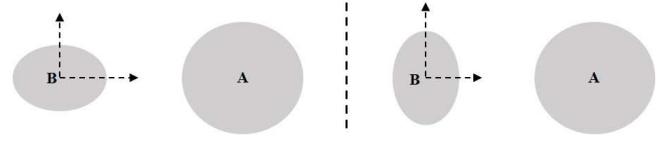}
			\caption{Configuration of equilibria.}
			\label{fig:Con}
		\end{figure}

	When the secondary enters the exact 1:1 spin-orbit resonance, the mutual orbit appears as circular and the secondary's long or short axis exactly points at the primary in one orbital period. Viewing the primary in the secondary's body-fixed frame, the primary {$A$} appears stationary in this frame, i.e. {$A$} is stationary, at the EPs of the secondary's long or short axis. In {$B$}'s body-fixed frame, the EPs are found by setting all velocities and accelerations to zero in Eqs.\ref{con:EOM2}. As a result, we have
		\begin{equation}
			\left\{
				\begin{array}{lr}
				\theta = 0 , \frac{\pi}{2} , \pi , \frac{3\pi}{2}\\
				0 = (\frac{K}{I_z^t})^2{r}^5-{r}^2-\frac{3}{2}(A_1+A_2\cos(2\theta))\\
						\end{array}
			\right.
			\label{con:EOM2e0}
		\end{equation}
		For a fixed value of $K$, the value of {$r$} can be solved from the second equation of Eq.\ref{con:EOM2e0}. Denote this value as $r_0$. Eq.\ref{con:EOM2e0} means that there are four EPs of Eq.\ref{con:EOM2}. Two lie on the long axis of {$B$}, and the other two lie on the short axis of $B$. Obviously, EPs with $\theta=0$ or $\theta=\pi$ are on {$B$}'s long axis and EPs with $\theta=\pi/2$ or $\theta=3\pi/2$ are on {$B$}'s short axis. Denote the distance from EP to secondary as $r_0$. Due to symmetry, there are two configurations of the exact 1:1 spin-orbit resonance, as shown in Fig.\ref{fig:Con}. The left configuration corresponds to the case of two EPs on {$B$}'s long axis, and the right configuration corresponds to the case of two EPs on {$B$}'s short axis. As a result, in the following we only focus on $\theta=0$ and $\theta=\pi/2$

		In this study, as an exmaple we will focus on the binary asteroid system (65803) Didymos. Some parameters of the system are given in Table.\ref{tab:Didymos_param} \citep{Michel2016}. According to Table \ref{tab:Didymos_param} and Eq.\ref{con:unit}, the length unit for this binary system is ${[L]}=975.6m$. As a result, the dimensionless mutual orbit is ${r} = 2.1525$ between the secondary and primary. Assuming that the primary is stationary at the EP exactly (i.e. ${r}_0 = {r}$), if the configuration is the left frame of Fig.\ref{fig:Con}, then the angular momentum is $K = 0.0132$, and, if the configuration is right frame of Fig.\ref{fig:Con}, then the angular momentum is $K = 0.0131$.

		\begin{table}
			\centering
			\caption{The dynamical and physical properties of Didymos.}
			\label{tab:Didymos_param}
			\begin{tabular}{lccr} 
				\hline
				properties & symbol & quatities\\
				\hline
				Mean diameter of the primary & $r_{A}$ & 780$m$ \\
				Mean diameter of the secondary & $\overline{a}_{B}$ & 163$m$\\
				Secondary(shape) elongation  & $a_{B}/b_{B}$ and $b_{B}/c_{B} $ & 1.2\\
				Bulk density of the primary  & $\rho$ & 2100$kg m^{-3}$\\
				Distance between component COMs  & $R$ & 2100$m$\\
				Secondary orbital period  & $P_{orb}$ & 11.92$hour$\\
				\hline
			\end{tabular}
		\end{table}

	\subsection{The parameter K}
\begin{figure}
			\includegraphics[width=\columnwidth]{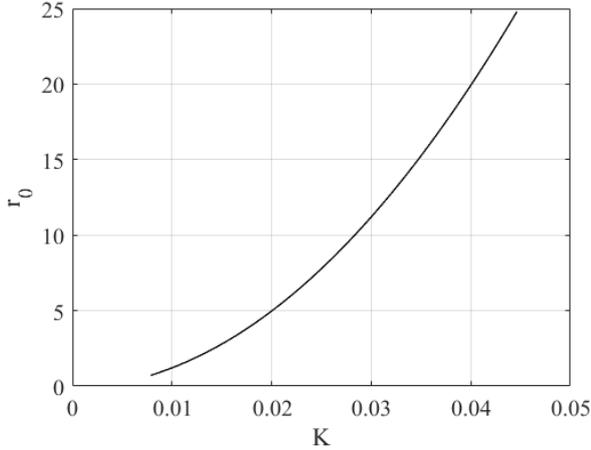}
			\caption{Position of the EPs in long axis changes with the total momentum $K$. It indecates that ${r}_0$ becomes lager as $K$ increase.}
			\label{fig:K_vs_S0}
		\end{figure}

		Eq.\ref{con:EOM2e0} means that the value of EP's position $r_0$ depends on the value of $K$. Taking the EP on {$B$}'s long axis as an example, Fig.\ref{fig:K_vs_S0} shows the relation between ${r}_0$ and $K$. Obviously, ${r}_0$ increases with $K$. As a result, $K$ is a parameter indicating the size of the synchronous system. Not all values of $K$ are feasible, because ${r}_0$ cannot be less than 1 in the units of Eq.\ref{con:unit}. Considering the condition ${r}_0 \ge 1$, the minimum value of $K$ can be obtained from Eqs.\ref{con:EOM2e0}, as shown in Eq.\ref{con:Kmin}. For the Didymos system, $K_{min} = 0.00913$ for the EP on the long axis ($\theta = 0$), and $K_{min} = 0.00908$ for the EP on the short axis ($\theta = \pi/2$).
		\begin{equation}
			K_{min} = \sqrt{\{1+\frac{3}{2}[A_1+A_2\cos(2\theta)]\}(m+I_z^{B})^2}
			\label{con:Kmin}
		\end{equation}

	\subsection{Stability of the EPs}
	\label{sec:EPslsa}

		Denote
		\begin{equation}
			X = ({r},\theta,\dot{{r}},\dot{\theta}) \notag
		\end{equation}
		The $2^{nd}$ order form of Eq. \ref{con:EOM2} can be rewritten in its $1^{st}$ order form as
		\begin{equation}
			\dot{X} = F(X,K)
		\label{con:Sim_EOM}
		\end{equation}
		where detailed form of $F(X,K)$ is a plain fact from Eq.\ref{con:EOM2}. The EP of the above equation is denoted as
		\begin{equation}
			X_0 = ({r}_0,\theta_0,0,0) \notag
		\end{equation}
		Expand Eq.\ref{con:Sim_EOM} around this EP, i.e., $X=X_0+\Delta X$, then we have
		\begin{equation}
			\Delta X = A\cdot \Delta X
		\end{equation}
		where
		\begin{equation}
				\textbf{A} = \left(\begin{array}{clclclclclcr}
				0	&	0	&	1	&	0\\
				0	&	0	&	0	&	1\\
				a_{31}	&	a_{32}	&	0	&	a_{34}\\
				a_{41}	&	a_{42}	&	a_{43}	&	a_{44}\\
				\end{array}\right) 
		\end{equation}
		Detailed expressions of $a_{ij}$ can be found in appendix \ref{sec:AppendA}. The characteristic equation for this matrix is simply. 
		\begin{equation}
				\lambda^4-(a_{31}+a_{42}+a_{34}a_{43})\lambda^2+a_{31}a_{42} = 0
		\label{con:egequation}
		\end{equation}
		Setting lowercase $s_i = \lambda^2$, roots of Eq.\ref{con:egequation} are,
		\begin{equation}
			\begin{array}{l}
				{s_1 = \frac{(a_{31}+a_{42}+a_{34}a_{43})}{2}+\frac{\sqrt{(a_{31}+a_{42}+a_{34}a_{43})^2-4a_{31}a_{42}}}{2}}\\
				{s_2 = \frac{(a_{31}+a_{42}+a_{34}a_{43})}{2}-\frac{\sqrt{(a_{31}+a_{42}+a_{34}a_{43})^2-4a_{31}a_{42}}}{2}}
			\end{array}
		\label{con:lambda}
		\end{equation}
		There are several cases for the stability of the EPs. If both $s_1$ and $s_2$ are real and both are smaller than zero, then the EPs are stable. If both $s_1$ and $s_2$ are real but one is larger than zero, then the EPs are hyperbolic unstable. If $s_1$ and $s_2$ are complex, then the EPs are complex unstable. 
		\begin{enumerate}
			\item $s_1>0,s_2>0$: saddle $\times$ saddle
			\item $s_2 \cdot s_2<0$: center $\times$ saddle
			\item $s_1<0,s_2<0$: center $\times$ center
			\item $s_1,s_2 \in$ Complex: complex unstable
		\end{enumerate}
		For the Didymos system in Table.\ref{tab:Didymos_param}, we have $s_1 = -0.0526;s_2 = -0.1014$ for the EP on {$B$}'s long axis, and  $s_1 = 0.0535;s_2 = -0.0999$ for the EP on $B$'s short axis. That means the EPs on $B$'s short axis are hyperbolic unstable, and EPs on {$B$}'s long axis are stable.

\begin{figure*}
			\begin{minipage}[t]{0.5\linewidth}
				\centering
				\includegraphics[width=\columnwidth]{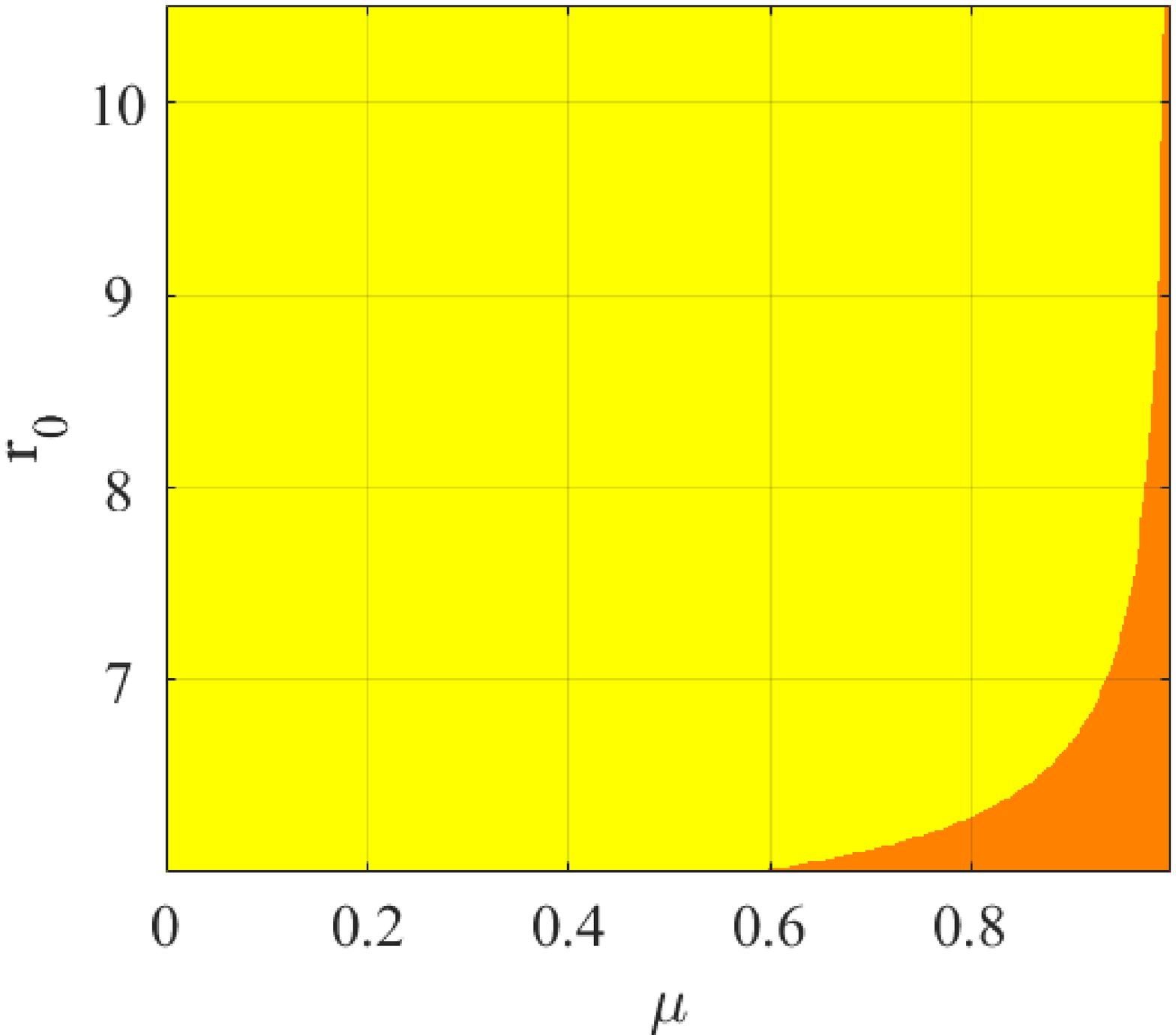}
			\end{minipage}%
			\begin{minipage}[t]{0.5\linewidth}
				\centering
				\includegraphics[width=\columnwidth]{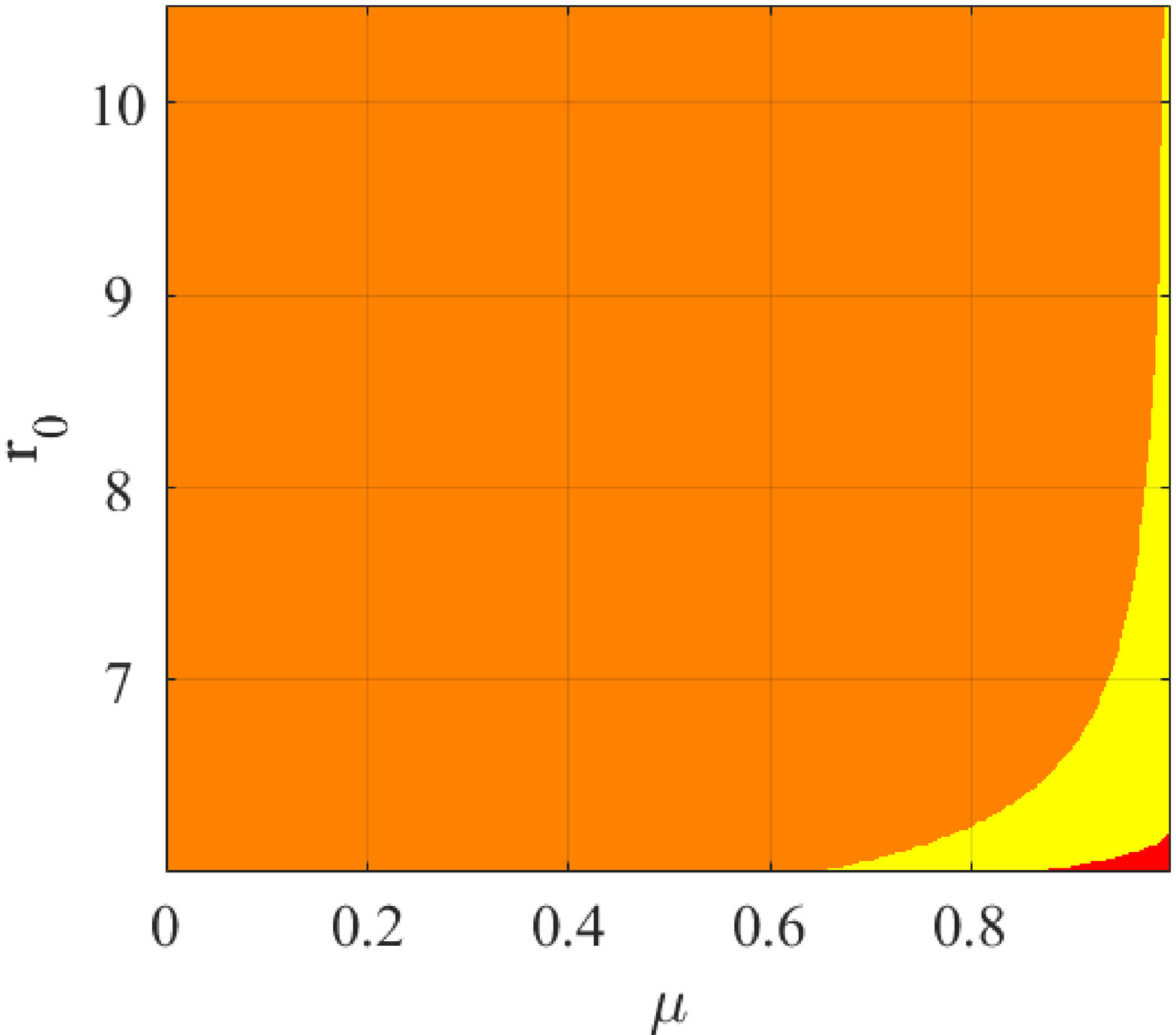}
			\end{minipage}
			\caption{Stability diagram of the EPs on the long axis (left) and the EPs on the short axis (right). For different values of ${r}_0$ and $\mu$, yellow is stable region, orange is hyperbolic unstable region and red is the complex unstable region.}
		\label{fig:Sta_Kmu}
		\end{figure*}

		We take a step further to study more general cases. We change the value of ${r}_0$ and $\mu$ to see how the stability of EPs change with these parameters. {Fig.\ref{fig:Sta_Kmu} shows the stability contour map w.r.t. the distance $r_0$ and the mass parameter $\mu$. The left frame is for the EPs on $B$'s long axis and the right frame is for the EPs on $B$'s short axis. Notice that ${r}_0$ indicates the distance between the EP and the secondary, and $\mu$ indicates the ratio of $B$'s mass to the total mass. For the EPs on $B$'s long axis, they are stable for small values of $\mu$. However, they may become unstable when $\mu$ is larger than 0.5, i.e., $B$ becomes the primary. For a fixed value of $\mu$, there is a limiting value $r_0$, smaller than which the EPs on the short axis becomes unstable. This phenomenon is already pointed out by previous studies \citep{Scheeres2009,Hou2017}. For Eps on the short axis, they are unstable for small values of $\mu$, i.e., a large primary with a small ellipsoid secondary. However, they can be stable if $\mu$ is large enough, i.e., the ellipsoid {$B$} in Fig.\ref{fig:Model} becomes the primary and the sphere {$A$} becomes the secondary. In the limiting case $\mu=1$, this is identical to the stability problem of EPs around a uniformly rotating ellipsoid studied by previous work \citep{Scheeres1994,Feng2017}.}  
		
\section{Families of Periodic Orbits around Equilibrium Points}
\label{sec:FOPO}
	According to \citep{Pravec2016}, {for binary asteroid systems discovered till now the $b_{B}/a_{B}$ value of the secondary is mostly in the interval $0.65$ to $1$, i.e., $a_B/b_B$ value is in the interval 1 to 1.53.} For a typical mutual distance of about 4 times the primary's radius \citep{Walsh2015} and a mass ratio of secondary to primary less than $0.2$, according to Fig.\ref{fig:Sta_Kmu} the EPs on the secondary's long axis are stable. Focusing on this stable equilibrium point, this section studies periodic orbits around it.

	\subsection{Solution to the linearized system}
	\label{sec:4.1}
\begin{figure}
			\includegraphics[width=\columnwidth]{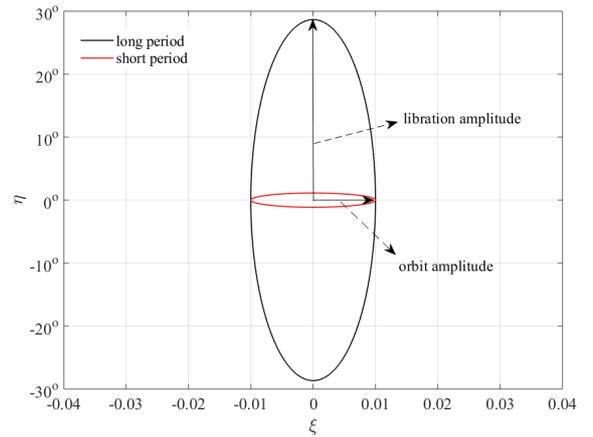}
			\caption{The configuration of long and short-period orbit in the linearized system. Note that the ordinate $\eta$ is not the $Y$ coordinate of the body-fixed frame, but the deviation of the angle $\theta$ from the EP (see the definition of $\eta$ at the beginning of section \ref{sec:4.1}). Same definition of orbit amplitude and libration amplitude also goes for the short-period orbit.}
			\label{fig:dtds}
		\end{figure}

		Let ${r} = {r}_0+\xi, \theta = \theta_0+\eta$, expand Eq.\ref{con:EOM2} around the EP, and only retain the linear terms, then we have Eq.\ref{con:Linearized_EOM}.  

		\begin{equation}
			\left\{
				\begin{array}{lr}
				\ddot{\xi}=a_{31}\xi+a_{34}\dot{\eta}\\
				\ddot{\eta}=a_{42}\eta+a_{43}\dot{\xi}\\
				\end{array}
			\right.
			\label{con:Linearized_EOM}
		\end{equation}
		For the EP at the secondary's long axis, we have $\theta_0$=0. Solution to the above equation is
		\begin{equation}
			\left\{
				\begin{array}{lr}
				\xi = C_1 \cos(\theta_1)+D_1\sin(\theta_1)+C_2\cos(\theta_2)+D_2\sin(\theta_2)\\
				\eta = \overline{C_{1}}\cos(\theta_1)+\overline{D_{1}}\sin(\theta_1)+\overline{C_{2}}\cos(\theta_2)+\overline{D_{2}}\sin(\theta_2)\\
							\end{array}
			\right.
			\label{con:Linearized_system}
		\end{equation}
		for which following relation holds
		\begin{equation}
				\overline{C}_{i} = \alpha_i D_i \quad \overline{D_{i}} = -\alpha_i C_i
		\end{equation}
		\begin{equation}
			\alpha_i = -\frac{a_{43} \omega_{i} }{a_{42}+\omega_{i}^{2}}=\frac{a_{31}+\omega_{i}^{2}}{a_{34} \omega_{i}}
		\label{con:alpha_i}
		\end{equation}
		in which $\omega_1=\sqrt{-s_1}$ and $\omega_2=\sqrt{-s_2}$ are two basic frequencies of the linearized system. Expressions of $s_1, s_2$ are given by Eq.\ref{con:lambda}. The constants of integration $C_1, C_2, D_1, D_2$ are functions of the initial conditions $\xi_0,\eta_0,\dot{\xi}_0,\dot{\eta}_0$, in the form of
		\begin{equation}
			C_{1} =\frac{\alpha_{2} \omega_{2} \xi_{0}+\dot{\eta}_{0}}{\alpha_{2} \omega_{2}-\alpha_{1} \omega_{1}} \quad \quad
			C_{2} =\frac{\alpha_{1} \omega_{1} \xi_{0}+\dot{\eta}_{0}}{\alpha_{1} \omega_{1}-\alpha_{2} \omega_{2}}\\
		\end{equation}
		\begin{equation}
			D_{1} =\frac{\omega_{2} \eta_{0}-\alpha_{2} \dot{\xi}_{0}}{\alpha_1\omega_{2}-\alpha_{2} \omega_{1}} \quad \quad
			D_{2} =\frac{\omega_{1} \eta_{0}-\alpha_{1} \dot{\xi}_{0}}{\alpha_2\omega_{1}-\alpha_{1} \omega_{2}}\\
		\end{equation}
		According to the above expressions, we know that by properly choosing the initial value, we can set $C_2=D_2=\overline{C}_2=\overline{D}_2=0$ and Eq.\ref{con:Linearized_system} is reduced to the form of
		\begin{equation}
			\left\{\begin{array}{l}
				{\xi=\xi_{0} \cos \theta_1+\frac{\eta_{0}}{\alpha_{1}} \sin \theta_1} \\
				{\eta=\eta_{0} \cos\theta_1-\alpha_{1} \xi_{0} \sin\theta_1}
			\end{array}\right.
			\label{con:NL}
		\end{equation}
		which is a periodic orbit with a period of $2\pi/\omega_1$. We can also set $C_1=D_1=\overline{C}_1=\overline{D}_1=0$, and Eq.\ref{con:Linearized_system} is reduced to the form of 
		\begin{equation}
			\left\{\begin{array}{l}
				{\xi=\xi_{0} \cos \theta_2+\frac{\eta_{0}}{\alpha_{2}} \sin \theta_2} \\
				{\eta=\eta_{0} \cos\theta_2-\alpha_{2} \xi_{0} \sin\theta_2}
			\end{array}\right.
			\label{con:NS}
		\end{equation}
		which also describes a periodic orbit, with a period of $2\pi/\omega_2$. According to the above studies, we know that there are two basic frequencies $\omega_1$ and $\omega_2$ for motions around the stable EP. Suppose $\omega_1\le \omega_2$, we call the periodic orbit given by Eq.\ref{con:NL} long-period orbit and the periodic orbit given by Eq.\ref{con:NS} short-period orbit. As an example, Fig.\ref{fig:dtds} shows the short- and the long-period orbit. We denote the amplitude of the period orbit as the distance between the EP and the right intersection point of the periodic orbit with the $\xi$ axis, and we denote the libration width as the absolute value of the $\eta$ coordinate of the uppermost/lowermost point in Fig.\ref{fig:dtds}. From this figure, it is obvious that the long-period orbit is more elongated than the short-period orbit, i.e., for same orbit amplitude, the libration amplitude of the long-period orbit is much larger than the short-period orbit. 

	\subsection{Families of long and short period orbits}
\begin{figure*}
			\begin{minipage}[t]{0.5\linewidth}
				\centering
				\includegraphics[width=\columnwidth]{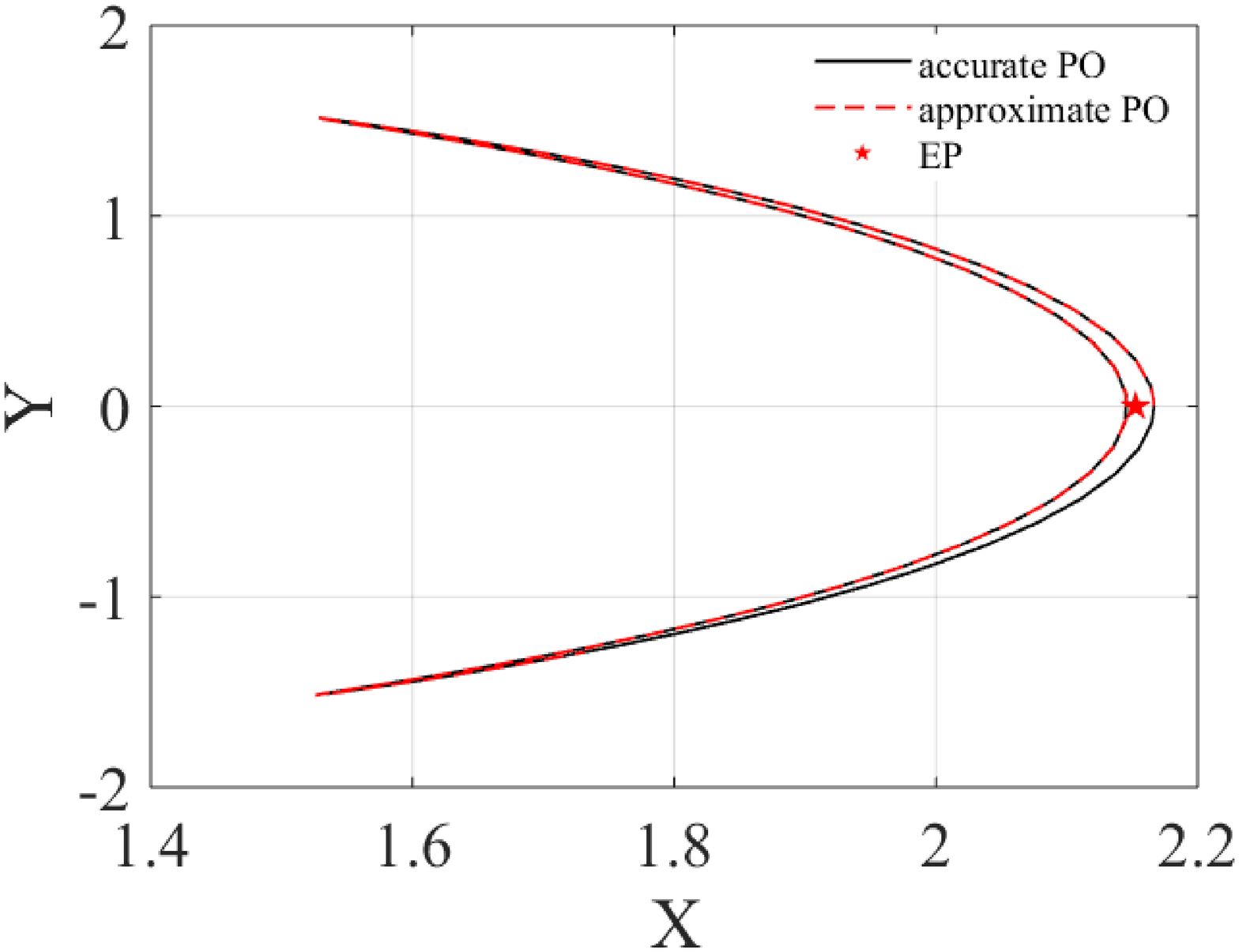}
			\end{minipage}%
			\begin{minipage}[t]{0.5\linewidth}
				\centering
				\includegraphics[width=\columnwidth]{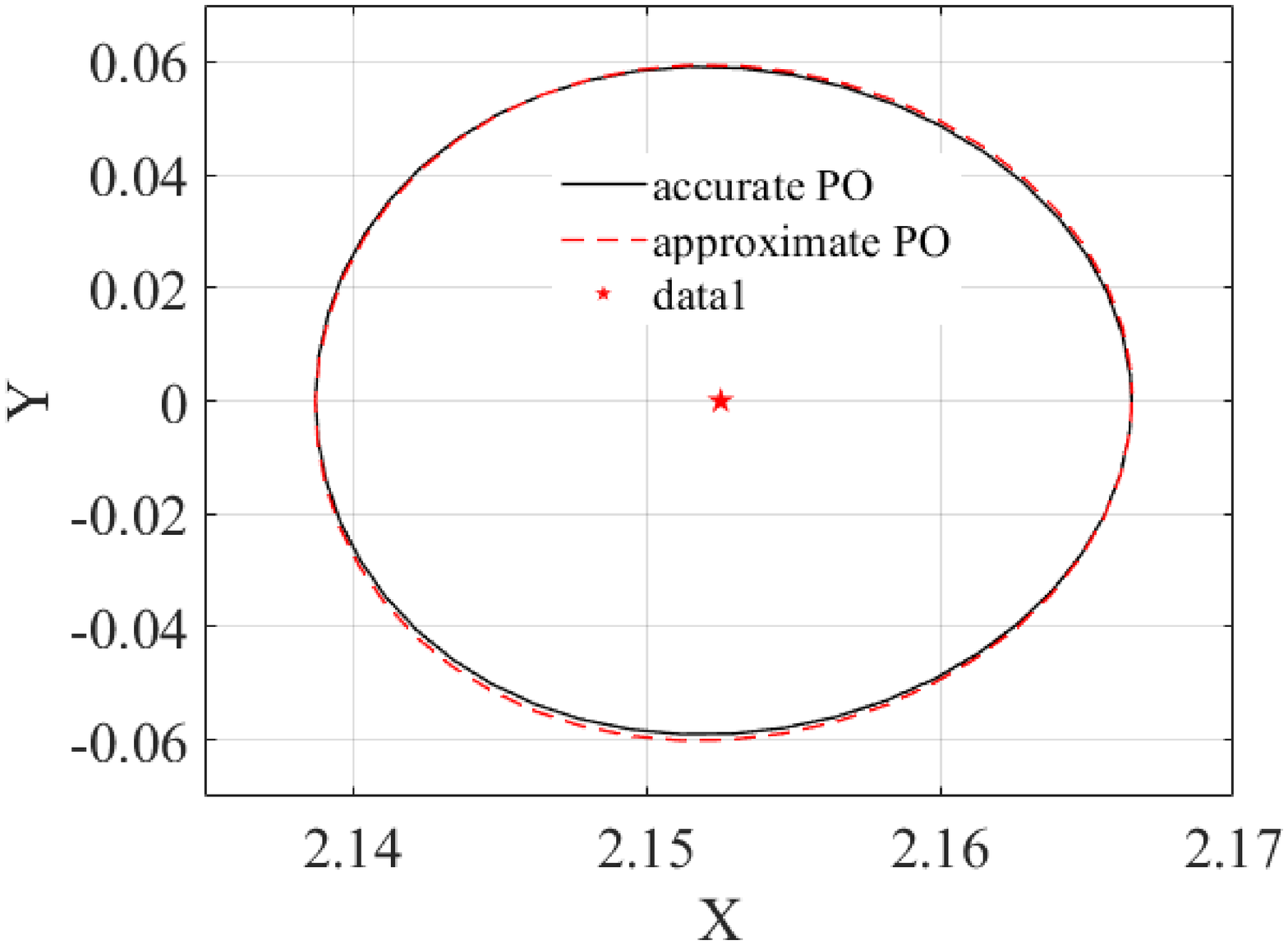}
			\end{minipage}
			\caption{One example long-period orbit (left) and one example short-period orbit (right) in $B$'s body-fixed frame, for which $K = 0.01$ and the orbit amplitude is 0.014. In each frame, the black curve is the accurate periodic orbit, and the red dashed curve is the linearized solution. The red star is the EP.}
			\label{fig:PO_LS}
		\end{figure*}

		According to Lyapunov's center theorem, under the assumption that $\omega_1$ and $\omega_2$ are incommensurate, there are two families of periodic orbits emanating from the stable EP. Starting from the linearized solution given by Eq.\ref{con:NL} and Eq.\ref{con:NS}, these two families of periodic orbits can be computed using the well-known predict-correct algorithm. Details of this technique are omitted here. Readers can refer to references (for example, \citet{Henrard1970,Doedel2007,Lei2018,Xin2016}). For large amplitude periodic orbit, one remark is that the linear solution obviously deviates from the true periodic orbit, as the example shown in Fig.\ref{fig:PO_LS}. For the short-period orbit shown in the right frame, the black curve is the true periodic orbit and the red dashed curve is the integrated orbit with initial conditions provided by the linear solution. The deviation between the two is observable but not so obvious. For the long-period orbit shown in the left frame, for the same orbit amplitude, the two orbits seem close to each other. However, notice that linear solution is far from being a closed curve. One remark is that the coordinate of Fig.\ref{fig:PO_LS} is the body-fixed frame of $B$, but not the one in Fig.\ref{fig:dtds}.

	\subsection{Genealogy and stability of families of periodic orbits}
	\label{sec:stability}
\begin{figure*}
			\begin{minipage}[t]{0.5\linewidth}
				\centering
				\includegraphics[width=\columnwidth]{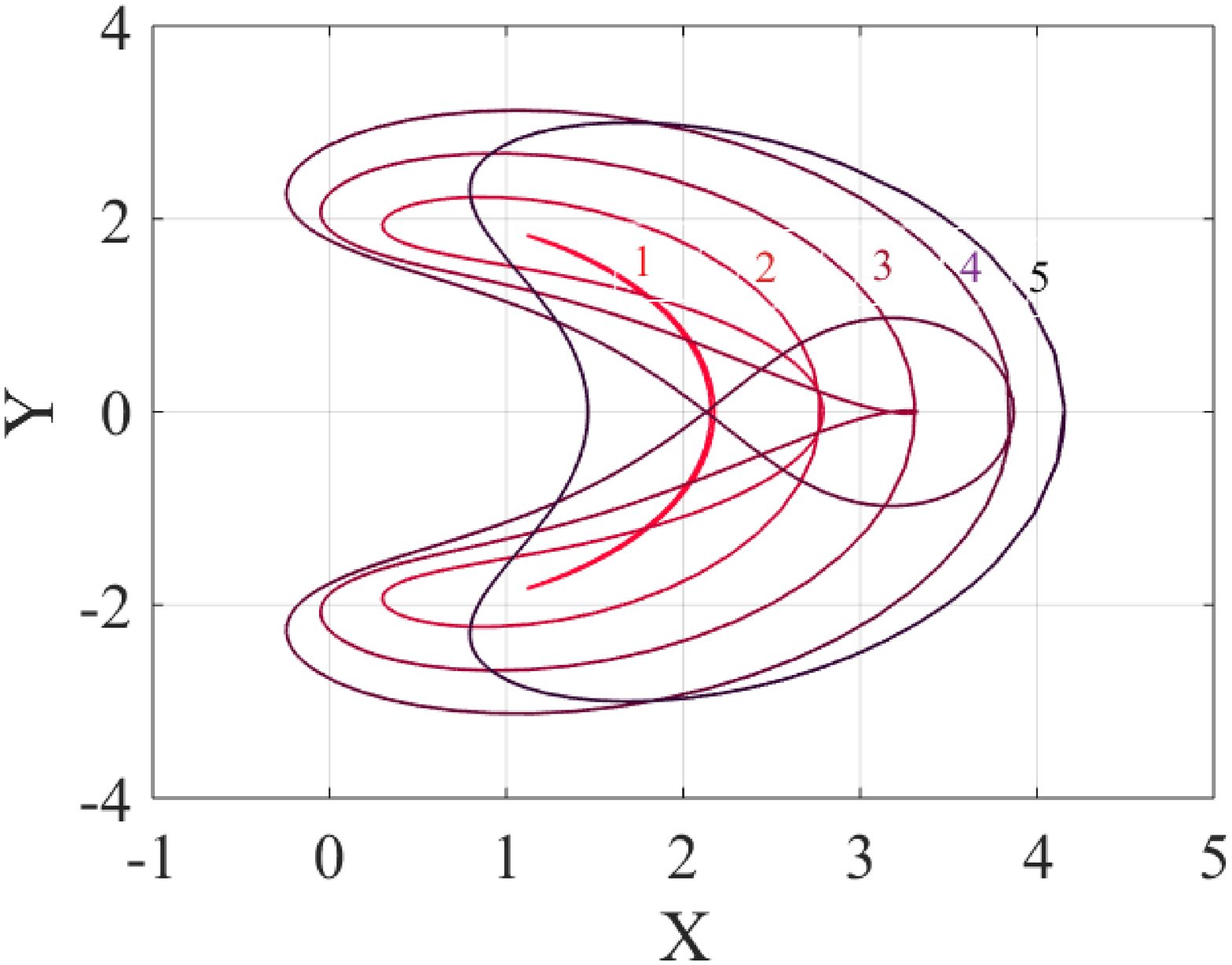}
			\end{minipage}%
			\begin{minipage}[t]{0.5\linewidth}
				\centering
				\includegraphics[width=\columnwidth]{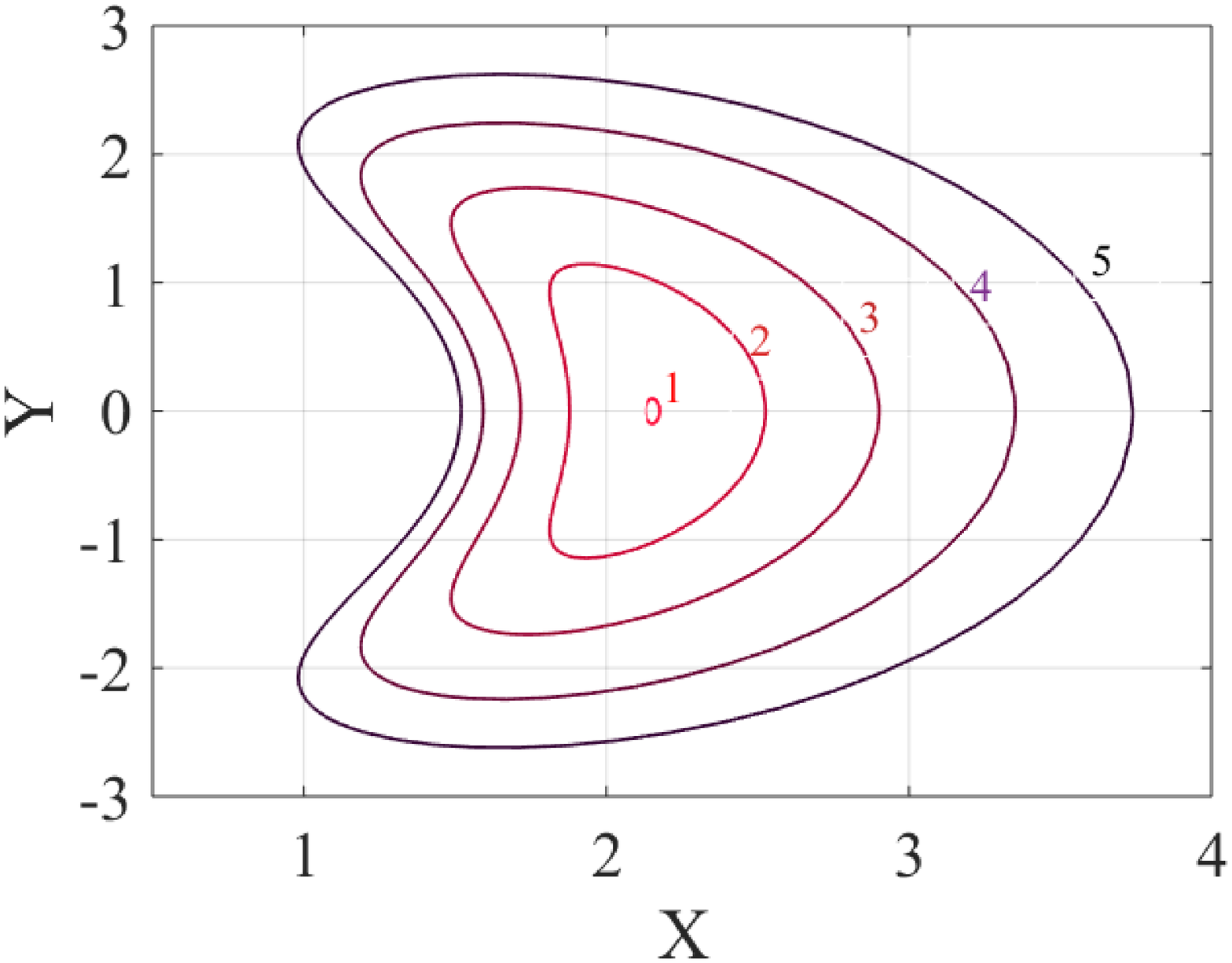}
			\end{minipage}
			\caption{Some example long-period orbits (left) and short-period orbits (right) in {$B$}'s body-fixed frame. The numbers indicate their positions in the ${r}-T$ curve shown in Fig.\ref{fig:PO_ST}. For the long-period orbits, with the period increasing, the orbit's shape first becomes more elongated, then twines itself, and finally terminates onto a short-period orbit traveling twice. For the short-period orbits, with the period increasing, the amplitude of the short-period orbits simply increases.}
			\label{fig:PO_family}
		\end{figure*}

\begin{figure*}
			\includegraphics[width=2\columnwidth]{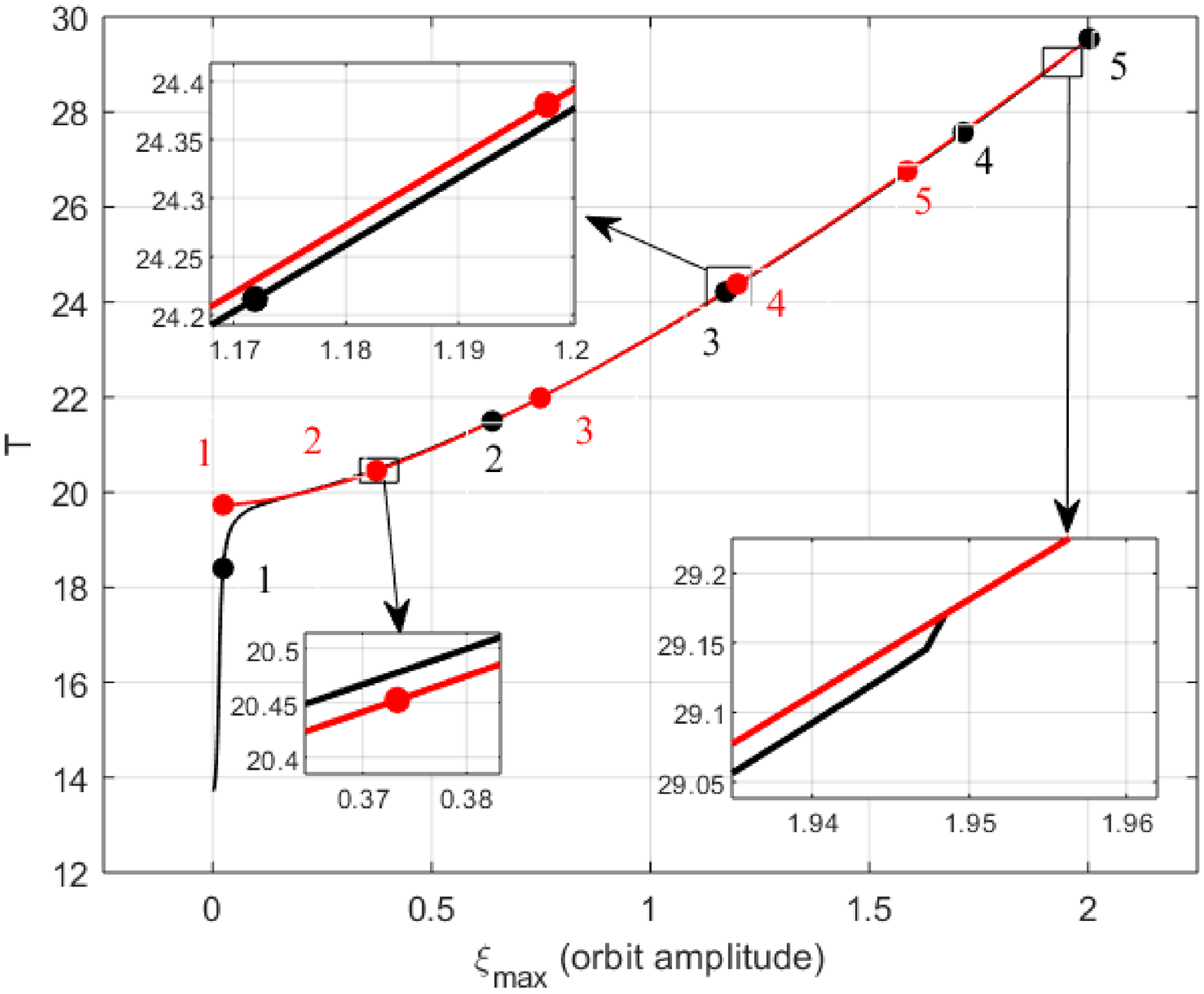}
			\caption{Characteristic curves of the long- and the short-period families. Black curve indicates the long-period orbits, and red curve is the short-period orbits with their period doubled.The black dots correspond to orbits in the left of Fig.\ref{fig:PO_family}, and the red dots correspond to orbits in the right of Fig.\ref{fig:PO_family}. The two curves intersect each other at the critical short-period orbit (see local magnification of this figure).}
			\label{fig:PO_ST}
		\end{figure*}

		Fig.\ref{fig:PO_family} shows some example orbits of the long-period family (left) and the short-period family (right). For the long-period family, with the period increasing, the long-period orbit's shape becomes more elongated, twines, and finally terminates onto a short-period orbit traveling twice. Fig.\ref{fig:PO_ST} shows the characteristic curves of these two families. The abscissa is the orbit amplitude and the ordinate is the period. The black dots in Fig.\ref{fig:PO_ST} correspond to orbits in the left of Fig.\ref{fig:PO_family}, and the red dots correspond to orbits in the right of Fig.\ref{fig:PO_family}. The fact that the long-period orbit terminates onto a short-period orbit traveling twice is obvious in this figure. 

		For the case shown in Fig.\ref{fig:PO_family} and Fig.\ref{fig:PO_ST}, $\omega_2/\omega_1 \in [1,2]$, and the long-period family terminates onto a short-period orbit traveling two times. By using different parameters for the binary system, our studies find that if $\omega_2/\omega_1 \in [k,k+1]$, the long-period family terminates onto a short-period orbit traveling $k+1$ times. We report this finding without giving any further details. This finding is interesting. It means that the global genealogy of the long- and the short-period families are same as that of the long- and the short-period families around triangular libration point of the circular restricted three-body problem \citep{Henrard2002,Hou2009On,Hou2009bridges} and also the same as that of the two periodic orbit families around the EP in the body-fixed frame of a single uniformly rotating asteroid \citep{Feng2017,Jiang2019}. 

\begin{figure}
			\includegraphics[width=\columnwidth]{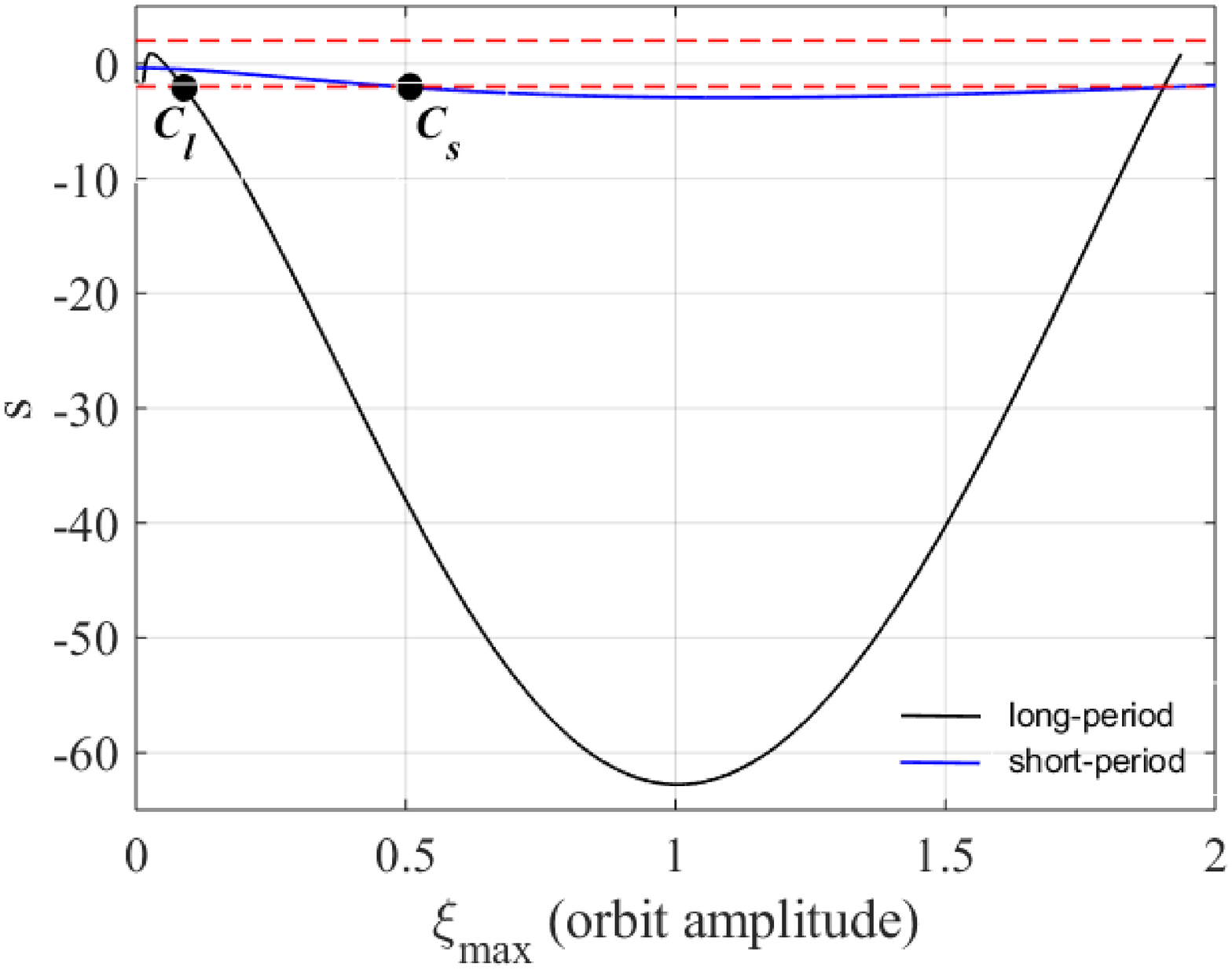}
			\caption{Stability curve of the long-period family and the short-period family. The ordinate is $s = trace(M)-2$ and the abscissa $\xi_{max}$ is the orbit amplitude. Black curve indicates the stability of the long-period family and blue curve is the short-period family. The red dashed lines are boundary of the stable region which ranges from -2 to +2. $C_l$ is the critical orbit of long-period family and $C_s$ is the critical orbit of short-period family.}
			\label{fig:PO_StrM}
		\end{figure}

\begin{figure}
			\includegraphics[width=\columnwidth]{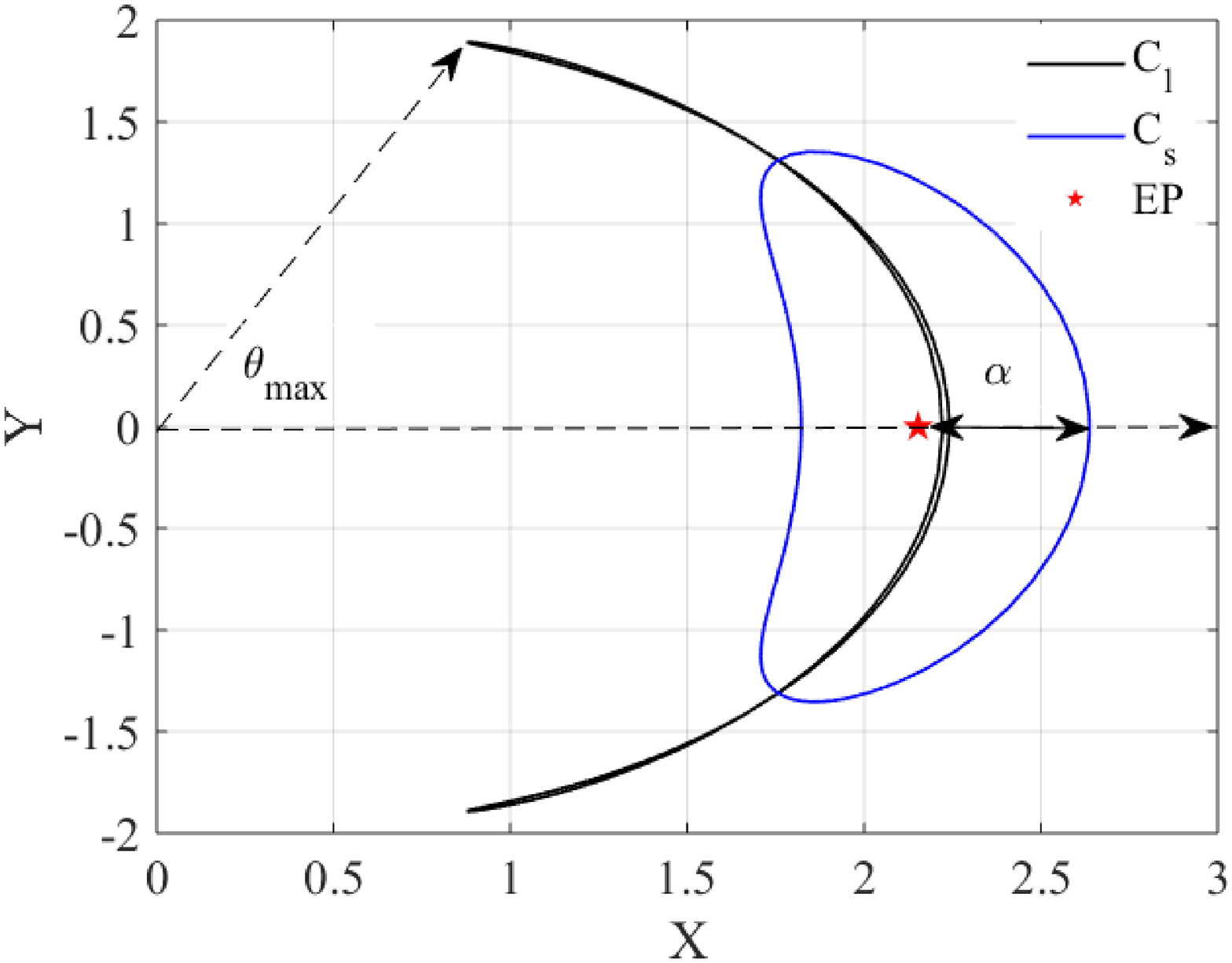}
			\caption{The critical orbits of the primary in the secondary's body-fixed fram for the Didymos system. The orbits $C_l$ and $C_s$ are already denoted in Fig.\ref{fig:PO_StrM}. $\theta_{max}$ is defined as the maximum libration angle. $\alpha$ is the maximum deviation of short-period orbit in $X$ axis. Red star is the EP.}
			\label{fig:Cls}
		\end{figure}

		Stability of periodic orbit is usually characterized by its Monodromy matrix $M$, which is the state transition matrix taken at one orbital period. For the Hamiltonian system, eigenvalues of the matrix $M$ usually appear in pairs \citep{Arnold1983}, i.e., if $\lambda$ is the eigenvalue of $M$, $\lambda^{-1}$ is also the eigenvalue of $M$. Since we are treating with an autonomous system, one pair of the eigenvalue is always 1. As a result, eigenvalues of the matrix $M$ for the planar long-period or the planar short-period orbit in our case are of the form ${1,1,\lambda,\lambda^{-1}}$. We define the stability index as $s=\lambda+\lambda^{-1}=trace(M)-2$. The periodic orbit is stable if $|s|<2$ and is unstable if $|s|>2$ \citep{Hou2008}. For the long-period and the short-period orbit studied in this subsection, the curve of the stability index w.r.t. the orbit amplitude is given in Fig.\ref{fig:PO_StrM}. The black curve is for the long-period family and the blue curve is for the short-period family. The two dashed lines indicate the region $-2\le s \le 2$. Judging from the figure, we know that both long-period orbits and short-period orbits are stable if their amplitude is small. With the amplitude increasing, the long-period orbit becomes unstable at the critical orbit $C_l$, and the short-period orbit becomes unstable at the critical orbit $C_s$, as illustrated in Fig.\ref{fig:PO_StrM}. The shapes of the two critical orbits $C_l$ and $C_s$ are given in Fig.\ref{fig:Cls}. Judging from Fig.\ref{fig:PO_StrM}, the long-period orbit is unstable when it larger than the orbit $C_l$, or the short-period orbit is unstable when it larger than the orbit $C_s$. In the following section, we will use these two critical orbits to study the maximum orbit eccentricity or the maximum libration ampltidue of the 1:1 spin-orbit resonance for binary asteroid systems.

\section{Stability Analysis of the 1:1 Spin-Orbit Resonance}
\label{sec:StaSOR}
	In the body-fixed frame of the secondary, in section \ref{sec:SOR} we have studied the EPs which correspond to exact 1:1 spin-orbit resonances of the secondary, and in section \ref{sec:FOPO} we have studied families of periodic orbits emanating from the stable EP. In this section, we interpretate these results in the physical space, trying to relate our findings with classical spin-orbit theory and some observation facts of the binary asteroid system.

	\subsection{Physical interpretation of two families of periodic orbits}
	\label{sec:5.1}

		This section relates the long- and the short-period family to the results of classical spin-orbit theory. We find that under the limiting case ($\mu \rightarrow 0$), our model reduces to the classical spin-orbit model. The long-period frequency $\omega_1$ is actually the free libration frequency of the classical spin-orbit model $\omega_{lib}$, and the short-period frequency $\omega_2$ is actually the frequency of the forced libration induced by the mutual orbit's eccentricity in the classical spin-orbit model, which is also the orbital frequency $n$. Further, we find that the amplitude of short-period orbit can be interpreted as the orbit eccentricity. The libration width of the short-period orbit is exactly the forced libration amplitude given in the classical spin-orbit model for $\mu \rightarrow 0$, but gradually deviates from it for large values of $\mu$, i.e., large secondary to primary mass ratio. This phenomenon is already pointed out by \citep{Naidu2015} in their numerical simulations. Here we provide an explanation to it.

		First, we make a very brief introduction to the classical spin-orbit theory. Readers can refer to \citep{Murray1999} for more details. The classical theory assumes an invariant Keplerian mutual orbit and only studies the secondary's rotational motion, i.e., the angle $\Theta$ and the mutual distance {$r$} in Fig.\ref{fig:Model} are only functions of time, but not variables of the system. For the planar problem, a well-known fact is that equations of the secondary's rotational motion follows

		\begin{equation}
			\ddot{\theta}_{B} = \frac{3}{2}\frac{B-A}{C}\frac{Gm_A}{{r}^3}\sin(2\theta)
		\label{con:CRF}
		\end{equation}
		where,

		\begin{equation}
			A = \frac{m_{B}}{5}(b_{B}^2+c_{B}^2), B = \frac{m_{B}}{5}(a_{B}^2+c_{B}^2),C = \frac{m_{B}}{5}(a_{B}^2+b_{B}^2)
		\end{equation}
		Introducing $\gamma = \theta_{B}-pM$, expanding Eq.\ref{con:CRF} w.r.t. the mean anomaly $M$, and averaging Eq.\ref{con:CRF} over $M$, we have \citep{Murray1999}

		\begin{equation}
			\ddot{\gamma} = -\frac{3}{2}\frac{B-A}{C}n^2H(p,e)\sin(2\gamma)
		\end{equation}
		where we assume $m_{B} \ll m_{A}$ and $n^2 = Gm_{A}/{r}^3$ . For the 1:1 spin-orbit resonance in this study, $p=1$ and

		\begin{equation}
			H(1,e) = 1-\frac{5}{2}e^2+\frac{13}{16}e^4 \approx 1
		\end{equation}
		As a result, the free libration frequency $\omega_{lib}$ is

		\begin{equation}
			\omega_{lib}^2 = 3\frac{B-A}{C}n^2 = 3\frac{a_A^2-b_A^2}{C}n^2 = 12\frac{mJ_{22}\alpha_B^2}{I_z^B{r}_0^3}
		\label{con:wlib}
		\end{equation}
		Besides the free libration, the orbit eccentricity causes forced libration of the secondary's rotation, with a frequency same as the orbit frequency $n$, and an amplitude of \footnote{Due to the coordinate used in this study, the forced libration is not the angle $\gamma$, but the angle $\Phi$ in Fig. 5.16 of \citep{Murray1999}}

		\begin{equation}
			A_{forced} = \frac{2n^2e}{n^2-\omega_{lib}^2}
		\label{con:Classical_A}
		\end{equation}

		Next, we come back to the model used in this work. For the limiting case $\mu \rightarrow 0$, according to the expressions given in appendix A, we have 
		
		\begin{equation}
			a_{31} = -\dot{\Theta}^2,a_{34} = 0,a_{42}=-\frac{12mJ_{22}\alpha_B^2}{I_z^B{r}_0^3},a_{43} = -\frac{2 \dot{\Theta}}{{r}_0} \notag
		\end{equation}
		According to Eq.\ref{con:lambda}, we have

		\begin{equation}
			\begin{array}{lr}
				{\omega_1^2 =-\lambda^2 = -a_{42} =\frac{12mJ_{22}\alpha_B^2}{I_z^B{r}^3}}\\
				{\omega_2^2 =-\lambda^2 = -a_{31} = \dot{\Theta}^2 = n^2}
			\end{array}
		\end{equation}
		Obviously, the short-period frequency $\omega_2$ is the orbital frequency $n$ in the limiting case. As for the long-period frequency $\omega_1$, paying attention that we are using non-dimensional units and noting the fact that

		\begin{equation}
			I_z^{B} = \frac{C}{[M]{[L]}^2},m = \frac{m_Am_B}{[M]^2},\alpha_{B} = \frac{a_{B}}{[L]},{r} = \frac{{L}}{[L]}
		\end{equation}
		it's easy to know that $\omega_1 = \omega_{lib}$. As a result, for the limiting case where the secondary's mass is negligible, the long-period component actually corresponds to the free libration of the 1:1 spin-orbit resonance in the classical model.

		Finally, we study the libration width of the short-period orbit in the limiting case. Rewrite the linearized short-period orbit given by Eq.\ref{con:NS} as

		\begin{equation}
			\begin{array}{l}
				{\xi =\alpha \cos(\theta_2+\phi)}\\
				{\eta =-\alpha_2\alpha \sin(\theta_2+\phi)}
			\end{array}
		\label{con:LE}
		\end{equation}
		where

		\begin{equation}
			\alpha = \frac{1}{\alpha_2}\sqrt{\alpha_2^2\xi_0^2+\eta_0^2},\quad \cos(\phi) = \frac{\xi_0}{C},\quad \sin(\phi)=-\frac{\eta_0}{\alpha_2C}
		\end{equation}
		According to Eq.\ref{con:alpha_i}, we have 

		\begin{equation}
			\alpha_2\alpha =-\frac{a_{43}\omega_2\alpha}{a_{42}+\omega_2^2}
		\label{con:Model_E}
		\end{equation}
		In the limiting case $a_{42} = -\omega_{lib}^2,a_{43} = -2n/{r}_0$ , so we have

		\begin{equation}
			\alpha_2\alpha =\frac{1}{{r}_0}\frac{2n^2\alpha}{n^2-\omega_{lib}^2}
		\end{equation}
		If we define the amplitude parameter $\alpha$ of the short-period orbit as
		
		\begin{equation}
			\alpha ={r}_0 \hat{\alpha}
		\label{con:alpha_hat}
		\end{equation}
		the libration width of the short-period orbit (see its definition in Fig.\ref{fig:dtds}) is

		\begin{equation}
			\alpha_2\alpha = \frac{2n^2\hat{\alpha}}{n^2-\omega_{lib}^2}
		\label{con:Model_R}
		\end{equation}
		Comparing Eq.\ref{con:Model_R} with Eq.\ref{con:Classical_A}, it’s reasonable to say that $\hat{\alpha}$ is the orbit eccentricity $e$, and the libration width of the short-period orbit in the limiting case given by Eq.\ref{con:Model_R} is exactly the amplitude of the forced libration in the classical spin-orbit model given by Eq.\ref{con:Classical_A}. 

\begin{figure}
			\includegraphics[width=\columnwidth]{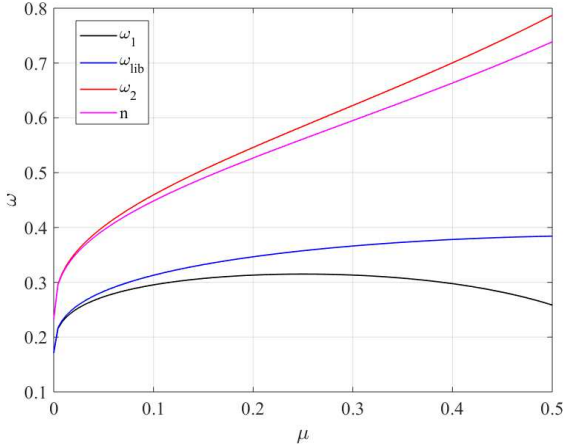}
			\caption{Curves of frequencies w.r.t. the mass parameter $\mu$. $\omega_1$ and $\omega_2$ are the long- and the short-period frequency. $\omega_{lib}$ is the frequency of the free libration computed by Eq.\ref{con:wlib}, and $n$ is the orbital frequency.}
			\label{fig:fre}
		\end{figure}

\begin{figure}
			\includegraphics[width=\columnwidth]{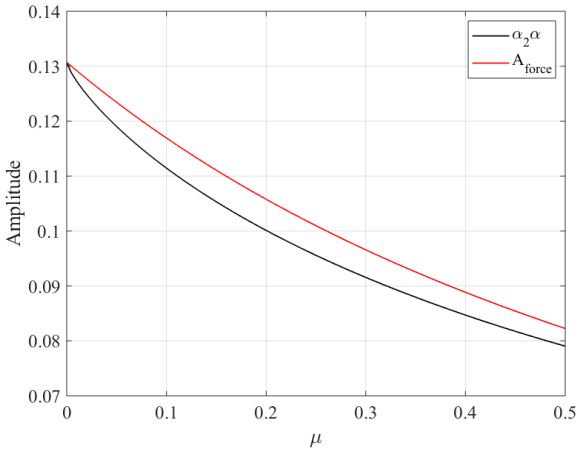}
			\caption{Black curve is the libration amplitude of the short-period orbit in our model which is computed by Eq.\ref{con:Model_E}. Red curve is the forced libration amplitude in the classical spin-orbit theory which is computed by Eq.\ref{con:Classical_A}. The two equals each other for the limiting case $\mu\rightarrow 0$, but obviously deviate from each other for large values of $\mu$.}
			\label{fig:amp}
		\end{figure}

		The above arguments tell us that our model can be reduced to the classical spin-orbit theory in the limiting case of $\mu\rightarrow 0$, by equating the long-period component with the free libration in the classical spin-orbit theory, and by equating the short-period component with the forced libration caused by the orbit eccentricity in the classical spin-orbit theory. However, for large values of $\mu$, due to the fact that the mutual orbit is far from being invariant, we can expect the obvious difference between the two models, as already pointed out by previous analytical work for spin-orbit resonances other than the 1:1 one \citep{Hou2017A}. In this study, we focus on the 1:1 spin-orbit resonance, by showing the difference between $\omega_1$ and $\omega_{lib}$, between $\omega_2$ and $n$, and between the short-period orbit's liberation amplitude with the forced libration amplitude by orbit eccentricity in classical theory. One example is given by Fig.\ref{fig:fre} and Fig.\ref{fig:amp}

		Using the Didymos shape parameters in Table.\ref{tab:Didymos_param}, we assume that the Didymos system enters the 1:1 exact spin-orbit resonance state. The distance between primary and secondary is fixed as $R = 2100m$, but we change the secondary's volume (i.e. secondary's mass $m_{B}$). As shown in Fig.\ref{fig:fre}, the short-period frequency $\omega_2$ equals the mutual orbit frequency $n$, and the long-period frequency $\omega_1$ equals the free libration frequency $\omega_{lib}$, when $\mu \rightarrow 0$. With the increase of $\mu$, difference between our model and the classical model becomes obvious.  

		Next, we compare the forced libration amplitude due to orbit eccentricity in the classical model which is computed by Eq.\ref{con:Classical_A}, with the libration amplitude of the short-period orbit in our model. The result is shown in Fig.\ref{fig:amp}. We use the linearized solution of the short-period orbit, so the libration amplitude of the short-period orbit is computed by Eq.\ref{con:Model_E},  with $\alpha$ computed by Eq.\ref{con:alpha_hat} where $\hat{\alpha}$ equals the orbit eccentricity $e$. In Fig.\ref{fig:amp}, we set $e=0.03$. As already being proved above, the two libration amplitudes equal each other for the limiting case $\mu \rightarrow 0$. However, according to Fig.\ref{fig:amp}, difference between the two is obvious for secondaries with non-negligible masses. This difference is already noticed by previous researchers in their numerical simulations \citep{Naidu2015,Pravec2016}. Here, we provide an explanation to the difference.
		
		Studies in this subsection indicate that our model can be reduced to the classical spin-orbit model in the limiting case of $\mu \rightarrow 0$, in the sense that the long-period component can be interpreted as the free libration and the short-period component can be interpreted as the forced libration caused by the orbit eccentricity. However, the two models show an obvious difference if the secondary's mass ratio $\mu$ is not negligible, which is often the case for binary asteroid systems. For such binary systems, influence from the secondary's non-spherical terms on the mutual orbit is not negligible, so the force model described by Eq.\ref{con:EOM3} or Eq.\ref{con:EOM2} is recommended.

	\subsection{Stability analysis based on periodic orbits}

\begin{figure}
			\includegraphics[width=\columnwidth]{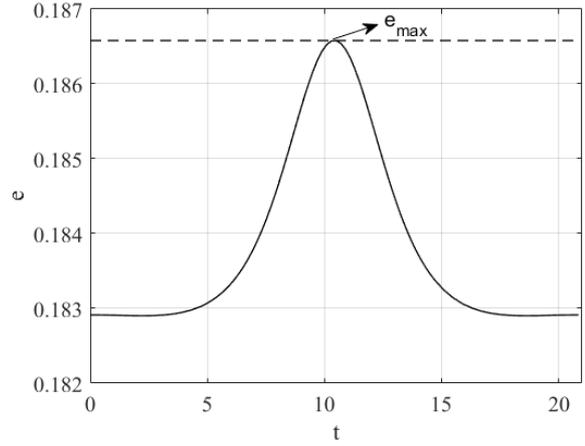}
			\caption{the value of orbital eccentricity of the critical short-period orbit during one period. Corresponding the definition of $\theta_{max}$ in Fig.\ref{fig:Cls}, the $e_{max}$ is defined as the maximum value during the critical short-period orbit.}
			\label{fig:emax}
		\end{figure}

		According to the arguments in the above subsection and Fig.\ref{fig:Cls}, we can safely say that the maximum libration amplitude $\theta_{max}$ of the critical long-period orbit $C_l$ is the maximum libration angle of the 1:1 spin-orbit resonance {(denoted in Fg.\ref{fig:Cls})}, which is usually smaller than $90^{\circ}$, and the maximum orbit eccentricity $e_{max}$ for the 1:1 spin-orbit resonance {(denoted in Fig.\ref{fig:emax})} is the maximum of eccentricity of the critical short-period orbit $C_s$. For a synchronous binary asteroid system, we can compute the long-period family and the short-period family and find out these critical orbits, as we did for the Didymos system. 

\begin{figure}
			\includegraphics[width=\columnwidth]{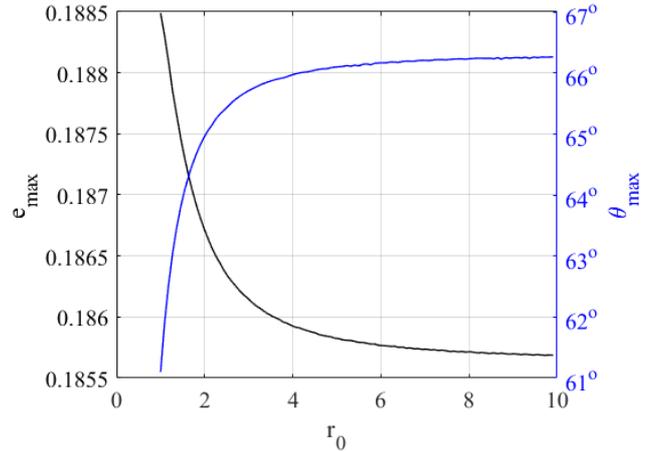}
			\caption{The upper limit of physical parameters in Didymos. Black line is the maximum of librational angles of long-period orbit. Blue line is the maximum value of orbital eccentricity of the critical short-period orbit.}
			\label{fig:ethteta}
		\end{figure}

		For a synchronous binary asteroid system, its mutual orbit is gradually altered by the so-called BYORP effect \citep{Cuk2005,Mcmahon2010} and the tidal torque. The mutual orbit may shrink, expand, or achieve a long-term balance \citep{Jacobson2011}, depending on the torque’s direction due to BYORP. As a result, for a specific binary asteroid system, it is interesting to study the maximum libration angle and maximum orbit eccentricity at different values of the mutual orbit distance. Taking the Didymos system as an example, Fig.\ref{fig:ethteta} shows the maximum libration angle $\theta_{max}$ and the maximum orbit eccentricity $e_{max}$ w.r.t. the mutual orbit distance ${r}_0$. From the figure, we know that the maximum libration angle/orbit eccentricity increases/decreases with the mutual orbit distance.

		If the BYORP torque works in the same direction as the tidal torque, the orbit gradually expands and the libration amplitude of the 1:1 spin-orbit resonance gradually increases due to the so-called adiabatic invariance \citep{Jacobson2014}. When it exceeds the maximum value $\theta_{max}$, the synchronous state is broken, forming the so-called wide asynchronous binary asteroid system \citep{Jacobson2014}. According to Eq. (93) and Eq. (95) of \citep{Mcmahon2010}, we have
		$$\frac{\dot{e}}{\dot{a}}=-\frac{e}{2a} \sim O(e)$$
		This means that the orbit eccentricity is also changing when the orbit is migrating inwards or outwards, but at a much lower speed if the orbit eccentricity is not large. As a result, it’s also possible that the orbit eccentricity exceeds the $e_{max}$ value before the libration angle exceeding the $\theta_{max}$ value, and the secondary’s synchronous state is broken. This mechanism may also provide a way to explain the origin of the asynchronous binary asteroid systems, especially for those with relatively large orbit eccentricities. Some further studies are necessary for this assumption and we will report our findings in a forthcoming paper.

		Results in Figs.\ref{fig:ethteta} put an upper limit on the orbit eccentricity for specific synchronous binary asteroid systems. Taking the binary systems with large orbit eccentricities listed in Table 1 of \citep{Pravec2016} as an example, Table \ref{tab:binaries} lists the upper limit of $e_{max}$ for these systems by us and those in \citep{Pravec2016}. Compare the $e_{max}$ values given by us and those in \citep{Pravec2016}, it seems that for some binary asteroid systems, their orbit eccentricity can be better constrained if we require the synchronous state to be stable.

		\begin{table*}
			\centering
			\caption{Maximum orbit eccentricity of some binary asteroid systems.$^a$}
			\label{tab:binaries}
			\begin{tabular}{p{3cm}p{1.5cm}p{1.5cm}p{1.5cm}p{1.5cm}p{1.5cm}p{1.5cm}p{1.5cm}} 
				\hline
				Binary system 		& $D_A(km)$ & $D_B(km)$ & $R(km)^b$ 	& $\mu^c$ & $a_{B}/b_{B}$ & $e_{max}$(Pravec) & $e_{max}$\\
				\hline
				(2131) Mayall 		& 8200 		& 2460 		& 19680 		& 0.0411 	& 1.31 		& 0.20 				& 0.1923\\
				(5481) Kiuchi 		& 3600 		& 1260 		& 7920 			& 0.0411 	& 1.30 	 	& 0.18 				& 0.1942\\
				(7088) Ishtar  		& 1050 		& 441 		& 2310 			& 0.0689	& 1.49 	 	& 0.16 				& 0.1794\\
				(76818) 2000 RG79  	& 2500 		& 850 		& 4250 			& 0.0378 	& 1.42 	 	& 0.13 				& 0.1856\\
				(17260) 2000 JQ58  	& 3300  	& 858 		& 5940 			& 0.0173 	& 1.54 	 	& 0.20 				& 0.1635\\
				(80218) 1999 VO123  & 880 		& 281 		& 2728 			& 0.0317 	& 1.52 	 	& 0.20 				&0.1647\\
				(5407) 1992 AX 		& 3700 		& 814 		& 6290 			& 0.0105 	& 1.52	 	& 0.11 				& 0.1643\\
				\hline
				\multicolumn{7}{l}{$^a$ The size and shape parameters are taken from Table 1 of \citep{Pravec2016}}\\
				\multicolumn{7}{l}{$^b$ The mutual distance is computed by taking a two-body relation from the masses and the orbital period, assuming a circular orbit.}\\
				\multicolumn{7}{l}{$^c$ For the two asteroids, their mass is computed by assuming $b_A=c_A$ and $b_B=c_B$.}\\
			\end{tabular}
		\end{table*}

\begin{figure}
			\includegraphics[width=\columnwidth]{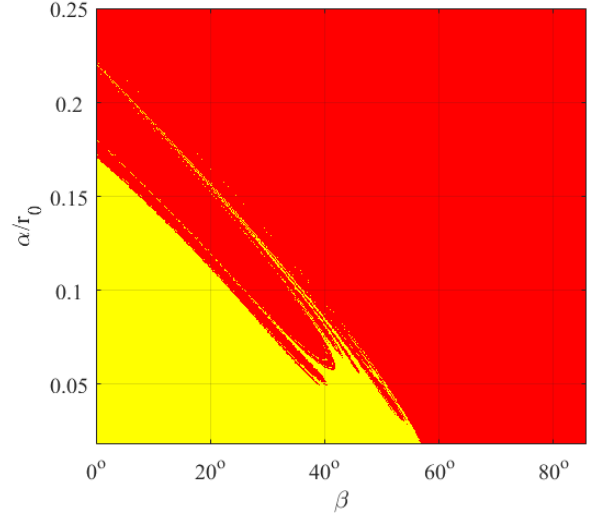}
			\caption{Stable region for the 1:1 synchronous state. Initial conditions of the orbits are provided by Eq.\ref{con:Linearized_system_alphabeta}. The abscissa $\beta$ is the amplitude of the long-period component, which according to Fig.\ref{fig:dtds} indicates the libration amplitude of the 1:1 synchronous state. The $\alpha$ indicates the amplitude of the short-period component, and the ordinate $\alpha/{r}_0$ which according to the above argument is an indicator of the orbit eccentricity. Yellow is the stable region while red is the unstable region.}
			\label{fig:AA}
		\end{figure}

		In reality, viewing the primary’s trajectory in the secondary’s body-fixed frame, it is not periodic but quasi-periodic, i.e., both the long-period and the short-period component exist. More accurate analytical description of the quasi-periodic motion is possible, by extending the linear solution by Eq.\ref{con:Linearized_system} to high orders, using the so-called Lindstedt-Poincaré technique \citep{Jorba1999,Hou2011} or other perturbation techniques. However, it is a tedious process which we try to avoid. Keeping to the linear solution, we carry out the following numerical experiment. First, we rewrite Eq.\ref{con:Linearized_system} as
		\begin{equation}
			\left\{
			\begin{array}{lr}
				\xi = \alpha\cos(\omega_2t+\phi_2)+\frac{\beta}{\alpha_1}\cos(\omega_1{t}+\phi_1)\\
				\eta = -\alpha_2\alpha\sin(\omega_2t+\phi_2)-\beta\sin(\omega_1{t}+\phi_1)\\
				\dot{\xi} = -\omega_2\alpha\sin(\omega_2t+\phi_2)-\frac{\beta\omega_1}{\alpha_1}\sin(\omega_1{t}+\phi_1)\\
				\dot{\eta} = -\omega_2\alpha_2\alpha\cos(\omega_2t+\phi_2)-\omega_1\beta\cos(\omega_1{t}+\phi_1)\\
			\end{array}
			\right.
		\label{con:Linearized_system_alphabeta}
		\end{equation}
		According to studies above, we know that $\alpha$ is the orbit amplitude of the short-period orbit and $\frac{\beta}{\alpha_1}$ is the orbit amplitude of the long-period orbit. According to Fig.\ref{fig:dtds} and studies in section \ref{sec:5.1}, we know that $\alpha/{r}_0$ is an indicator of the orbit eccentricity, and $\beta$ is an indicator of the libration amplitude. By choosing different values of $\alpha$ and $\beta$, we actually choose different trajectories. Integrate these trajectories for some time $T_{int}$. If the synchronous state is broken within this time, it means that the combination of $\alpha$ and $\beta$ is not possible for the synchronous state. If the synchronous state is preserved after this time, it means that the combination of $\alpha$ and $\beta$ corresponds to a stable synchronous state, at least within the integration time $T_{int}$. Taking the Didymos system as an example, Fig.\ref{fig:AA} shows the results of the numerical experiment for which $T_{int}=150$. 

		Judging from Fig.\ref{fig:AA}, an obvious feature is an anti-correlation between the orbit eccentricity and the libration amplitude of the synchronous state. That is, for a fixed value of the libration amplitude, there is a critical value of the orbit eccentricity $e_c$ beyond which the 1:1 synchronous state is unstable. The $e_c$ value decreases with increasing libration amplitude $\beta$. As a result, the larger the libration amplitude is, the smaller the possible orbit eccentricity for the stable 1:1 synchronous state is. This may help researchers understand the fact that most observed synchronous binary asteroid systems have small orbit eccentricities \citep{Pravec2016}, because the orbit eccentricity cannot be larger than $e_{max}$, which according to Fig.\ref{fig:ethteta} and Table \ref{tab:binaries}, is usually not large.

\section{Discussion}
\label{sec:Dis}
	The current study neglects the non-spherical terms of the primary. A natural question is what happens when the non-spherical terms of the primary are included. In this case, we have to simultaneously consider the primary’s rotation, and the model described in Fig.\ref{fig:Model} is changed to the one in the following Fig.\ref{fig:ModelAB}. Another angle $\phi$ is introduced to describe the relative geometry of the primary w.r.t. the secondary. In this case, equations of motion are changed from Eq.\ref{con:EOM3} to the form of \citep{Hou2017A,Hou2017}
	\begin{figure}
		\includegraphics[width=\columnwidth]{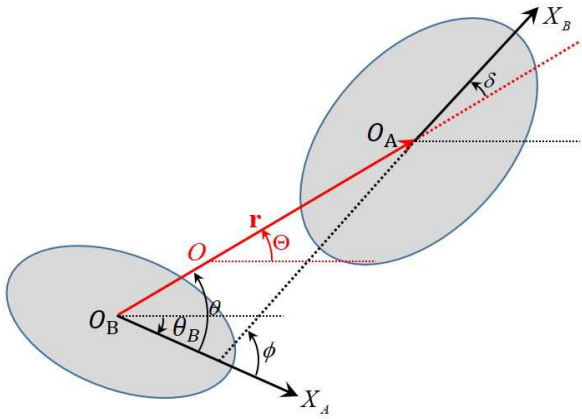}
		\caption{Relative geometry of the planar two-ellipsoid model for the binary asteroid system}
		\label{fig:ModelAB}
	\end{figure}

	\begin{equation}
		\left\{
			\begin{array}{lr}
				{\ddot{r} = r(\dot{\theta}+\dot{\theta}_B)^2-\frac{1}{r^2}-\frac{3}{2r^4}(A_1+A_2cos(2\theta)+A_3cos(2\delta))}\\
				{\ddot{\theta} = -2\frac{\dot{r}}{r}(\dot{\theta}+\dot{\theta}_B)-\frac{1}{r^5}(A_2sin(2\theta)+A_3sin(2\delta))-\frac{mA_2}{I_z^B}\frac{sin(2\theta)}{r^3}}\\
				{\ddot{\phi} = \frac{mA_3}{I_z^B}\frac{sin(2\delta)}{r^3}-\frac{mA_2}{I_z^B}\frac{sin(2\theta)}{r^3}}\\
				{\ddot{\theta}_B = \frac{mA_2}{I_z^B}\frac{sin(2\theta)}{r^3}}
			\end{array}
		\right.
	\label{con:EOMAB}
	\end{equation}
	where the coefficient $A_2$ is same as the one in Eq.\ref{con:EOM3}, and the coefficient $A_1$ is updated as
	\begin{equation}
		A_1 = \alpha_A^2J_2^A+\alpha_B^2J_2^B
	\label{con:A1}
	\end{equation}
	where $\alpha_B = \overline{a}_B/{[L]}$, and the coefficient $A_3$ is defined as
	\begin{equation}
		A_3 = 6\alpha_B^2J_{22}^B
	\label{con:A3}
	\end{equation}
	Usually, the introduction of the primary’s non-spherical terms also brings the frequency of the primary’s rotational motion to the system. If the primary’s rotational motion is not synchronous with its orbital motion, the exact synchronous state of the secondary no longer appears as an EP in the body-fixed frame of the secondary, but as a periodic orbit forced by the primary’s rotation. In literature, we call this forced motion as dynamical substitute\citep{Chappaz2015}, and the periodic orbits studied in the current work become quasi-periodic orbits around the periodic dynamical substitute.

	Nevertheless, the amplitude of the dynamical substitute is usually small. For synchronous binary asteroid systems discovered till now, the primary usually rotates much faster than the orbital motion, so we can simply average Eq.\ref{con:EOMAB} by simply neglecting the short-period terms for the 1:1 spin-orbit resonance. The equations of motion after average are
	\begin{equation}
		\left\{
		\begin{array}{lr}
			\ddot{r} = r(\dot{\theta}+\dot{\theta}_B)^2-\frac{1}{r^2}-\frac{3}{2r^4}[A_1+A_2\cos(2\theta)]\\
			\ddot{\theta} = -\frac{A_2}{r^5}\sin(2\theta)-2\frac{\dot{r}}{r}(\dot{\theta}+\dot{\theta}_B)-\frac{mA_2}{I_z^B}\frac{\sin(2\theta)}{r^3}\\
			\ddot{\theta}_B = \frac{mA_2}{I_z^Br^3}\sin(2\theta)\\
		\end{array}
		\right.
	\label{con:EOMAB3}
	\end{equation}
	which has an exactly same form as Eq.\ref{con:EOM3} but only the term $A_1$ differs from the one in Eq.\ref{con:EOM3}, in the sense that the extra term $\alpha_AJ_2^A$ appears (see Eq.\ref{con:A1A2} and Eq.\ref{con:A1}). Methods for the above studies can be also applied to Eq.\ref{con:A1}. Considering the fact that the term $\alpha_AJ_2^A$ is usually larger than the term $\alpha_BJ_2^B$ due to the fact that $A$ is larger than $B$, consideration of primary's non-spherical terms will introduce some modifications to the stability of the binary asteroid system. We will report these findings for the averaged system described by Eq.\ref{con:EOMAB3} in a forthcoming paper, along with the periodic contributions from the primary's rotational motion which also influence the stability of the binary asteroid system. Also, we will study the thermal effects which are neglected in the current study.

\section{Conclution}
\label{sec:Con}

	By using the simple planar full two-body problem and only considering the mutual gravity between the two bodies, the 1:1 spin-orbit resonance of the secondary in a synchronous binary asteroid system is studied in this paper. There are two differences of the current study from previous work. First, the orbital motion and the rotational motion are simultaneously considered. Due to the close mutual distance and highly non-spherical shape of the secondary, the model of considering a time-varying mutual orbit is reasonable, and shows obvious difference from the conventional model which assumes an invariant mutual orbit if the secondary's mass is non-negligible. Second, the approach of periodic orbits is taken in this study, different from the usual Hamiltonian approach. 

	Throughout the work, the 1:1 spin-orbit resonance is studied in the secondary's body-fixed frame, using a model composed of a sphere primary and an ellipsoidal secondary. By using the angular momentum $K$, the system is reduced to a dynamical system of 2-DOF. In the secondary's body-fixed frame, exact 1:1 spin-orbit resonance appears as EPs, and general 1:1 spin-orbit resonance orbits appear as quasi-periodic orbits around the EPs, with two basic frequencies. By studying the EPs and the periodic families, following findings are achieved:

	(1) For an ellipsoidal secondary, the EPs on the secondary's long axis are stable while the EPs on the short axis are unstable. For stable EPs, there are two families of periodic orbits generating from them. One is the long-period family and the other is the short-period family. Genealogy between the two families is the same as that of the two planar families around triangular libration point of the circular restricted three-body problem, and the same as that of the planar families around the EP in the body-fixed frame of a uniformly rotating asteroid.

	(2) In both families, there is a critical orbit larger than which the periodic orbit becomes unstable. For the critical orbit $C_l$ in the long-period family, its libration amplitude is the maximum libraiton amplitude of the 1:1 spin-orbit resonance, and for the critical orbit $C_s$ in the short-period family, its orbit amplitude is the maximum orbit eccentricity of the 1:1 spin-orbit resonance.

	(3) In the limiting case $\mu\rightarrow 0$, results in the classical spin-orbit model are recovered, by equating the long-period orbit of our model with the free libration of the classical model, and by equating the short-period orbit of our model with the forced libration of the classical model. However, the two models show obvious difference for secondaries with non-negligible masses.

	(4) We find the anti-correlation between the orbit eccentricity and the libration amplitude. For a fixed value of the libration amplitude, there is a critical value $e_c$ beyond which the synchronous state is unstable. This $e_c$ value decreases with increasing libration amplitude. That is, the larger the libraiton amplitude is, the smaller the orbit eccentricity is.

\section*{Acknowledgements}

This work is supported by the National Natural Science Foundation of China(11773017, 11703013, 11673072).




\bibliographystyle{mnras}
\bibliography{example} 

\begin{thebibliography}{}
\makeatletter
\relax
\def\mn@urlcharsother{\let\do\@makeother \do\$\do\&\do\#\do\^\do\_\do\%\do\~}
\def\mn@doi{\begingroup\mn@urlcharsother \@ifnextchar [ {\mn@doi@}
  {\mn@doi@[]}}
\def\mn@doi@[#1]#2{\def\@tempa{#1}\ifx\@tempa\@empty \href
  {http://dx.doi.org/#2} {doi:#2}\else \href {http://dx.doi.org/#2} {#1}\fi
  \endgroup}
\def\mn@eprint#1#2{\mn@eprint@#1:#2::\@nil}
\def\mn@eprint@arXiv#1{\href {http://arxiv.org/abs/#1} {{\tt arXiv:#1}}}
\def\mn@eprint@dblp#1{\href {http://dblp.uni-trier.de/rec/bibtex/#1.xml}
  {dblp:#1}}
\def\mn@eprint@#1:#2:#3:#4\@nil{\def\@tempa {#1}\def\@tempb {#2}\def\@tempc
  {#3}\ifx \@tempc \@empty \let \@tempc \@tempb \let \@tempb \@tempa \fi \ifx
  \@tempb \@empty \def\@tempb {arXiv}\fi \@ifundefined
  {mn@eprint@\@tempb}{\@tempb:\@tempc}{\expandafter \expandafter \csname
  mn@eprint@\@tempb\endcsname \expandafter{\@tempc}}}

\bibitem[\protect\citeauthoryear{Arnold \& V.}{Arnold \& V.}{1983}]{Arnold1983}
Arnold V. I.,  1983, Advances in Mathematics, 49, 106

\bibitem[\protect\citeauthoryear{Balmino}{Balmino}{1994}]{Balmino1994}
Balmino G.,  1994, \mn@doi [CeMDA] {Doi 10.1007/Bf00691901}, 60, 331

\bibitem[\protect\citeauthoryear{Bellerose \& Scheeres}{Bellerose \&
  Scheeres}{2008}]{Bellerose2008}
Bellerose J.,  Scheeres D.~J.,  2008, \mn@doi [CeMDA]
  {10.1007/s10569-007-9108-3}, 100, 63

\bibitem[\protect\citeauthoryear{Broucke}{Broucke}{1969}]{Broucke1969}
Broucke R.,  1969, AIAA J, 7, 1003

\bibitem[\protect\citeauthoryear{{Celletti}}{{Celletti}}{1990a}]{Celletti1990a}
{Celletti} A.,  1990a, \mn@doi [\zap] {10.1007/BF00945107}, \href
  {https://ui.adsabs.harvard.edu/abs/1990ZaMP...41..174C} {41, 174}

\bibitem[\protect\citeauthoryear{Celletti}{Celletti}{1990b}]{Celletti1990b}
Celletti A.,  1990b, \zap, 41, 453

\bibitem[\protect\citeauthoryear{Chappaz \& Howell}{Chappaz \&
  Howell}{2015}]{Chappaz2015}
Chappaz L.,  Howell K.~C.,  2015, CeMDA, 123, 123

\bibitem[\protect\citeauthoryear{Doedel, Romanov, Paffenroth, Keller  \&
  Dichmann}{Doedel et~al.}{2007}]{Doedel2007}
Doedel E.~J.,  Romanov V.~A.,  Paffenroth R.~C.,  Keller H.~B.,   Dichmann D.
  J. e.~a.,  2007, IJBC, 17, 2625

\bibitem[\protect\citeauthoryear{Fahnestock \& Scheeres}{Fahnestock \&
  Scheeres}{2006}]{Fahnestock2006}
Fahnestock E.~G.,  Scheeres D.~J.,  2006, \mn@doi [CeMDA]
  {10.1007/s10569-006-9045-6}, 96, 317

\bibitem[\protect\citeauthoryear{Feng \& Hou}{Feng \& Hou}{2017}]{Feng2017}
Feng J.,  Hou X.,  2017, \aj, 154, 21

\bibitem[\protect\citeauthoryear{Ferrari, Lavagna  \& Howell}{Ferrari
  et~al.}{2016}]{Ferrari2016}
Ferrari F.,  Lavagna M.,   Howell K.~C.,  2016, \mn@doi [CeMDA]
  {10.1007/s10569-016-9688-x}, 125, 413

\bibitem[\protect\citeauthoryear{Gabern, Koon, Marsden  \& Scheeres}{Gabern
  et~al.}{2006}]{Gabern2006}
Gabern F.,  Koon W.~S.,  Marsden J.~E.,   Scheeres D.~J.,  2006, SIADS, 5, 252

\bibitem[\protect\citeauthoryear{Goldreich \& Peale}{Goldreich \&
  Peale}{1966}]{Goldreich1966}
Goldreich P.,  Peale S.~J.,  1966, \nat, 209, 1078

\bibitem[\protect\citeauthoryear{Hadjidemetriou}{Hadjidemetriou}{1975}]{Hadjidemetriou1975}
Hadjidemetriou J.~D.,  1975, \mn@doi [Celestial mechanics]
  {10.1007/bf01230209}, 12, 155

\bibitem[\protect\citeauthoryear{Henrard}{Henrard}{1970}]{Henrard1970}
Henrard J.,  1970, \aap, 5, 45

\bibitem[\protect\citeauthoryear{Henrard}{Henrard}{2002}]{Henrard2002}
Henrard J.,  2002, CeMDA, 83, 291

\bibitem[\protect\citeauthoryear{Hirabayashi et~al.,}{Hirabayashi
  et~al.}{2019}]{Hirabayashi2019}
Hirabayashi M.,  et~al., 2019, \mn@doi [ASR]
  {https://doi.org/10.1016/j.asr.2018.12.041}, 63, 2515

\bibitem[\protect\citeauthoryear{Hou}{Hou}{2009}]{Hou2009On}
Hou X.~Y.,  2009, RAA, 9, 494

\bibitem[\protect\citeauthoryear{Hou \& Liu}{Hou \& Liu}{2008}]{Hou2008}
Hou X.~Y.,  Liu L.,  2008, CeMDA, 101, 309

\bibitem[\protect\citeauthoryear{Hou \& Liu}{Hou \& Liu}{2009}]{Hou2009bridges}
Hou X.~Y.,  Liu L.,  2009, CeMDA, 104, 241

\bibitem[\protect\citeauthoryear{Hou \& Liu}{Hou \& Liu}{2011}]{Hou2011}
Hou X.~Y.,  Liu L.,  2011, \mnras, 415, 3552

\bibitem[\protect\citeauthoryear{Hou \& Xin}{Hou \& Xin}{2017}]{Hou2017A}
Hou X.,  Xin X.,  2017, \mn@doi [Astrodynamics] {10.1007/s42064-017-0010-9}, 2,
  39

\bibitem[\protect\citeauthoryear{Hou, Scheeres  \& Xin}{Hou
  et~al.}{2017}]{Hou2017}
Hou X.,  Scheeres D.~J.,   Xin X.,  2017, \mn@doi [CeMDA]
  {10.1007/s10569-016-9731-y}, 127, 369

\bibitem[\protect\citeauthoryear{Jacobson \& Scheeres}{Jacobson \&
  Scheeres}{2011}]{Jacobson2011}
Jacobson S.~A.,  Scheeres D.~J.,  2011, \mn@doi [The Astrophysical Journal]
  {10.1088/2041-8205/736/1/l19}, 736, L19

\bibitem[\protect\citeauthoryear{Jacobson, Scheeres  \& McMahon}{Jacobson
  et~al.}{2014}]{Jacobson2014}
Jacobson S.~A.,  Scheeres D.~J.,   McMahon J.,  2014, \mn@doi [\apj] {Artn 60
  10.1088/0004-637x/780/1/60}, 780

\bibitem[\protect\citeauthoryear{Jiang \& Baoyin}{Jiang \&
  Baoyin}{2019}]{Jiang2019}
Jiang Y.,  Baoyin H.,  2019, \mn@doi [Results in Physics]
  {https://doi.org/10.1016/j.rinp.2018.11.049}, 12, 368

\bibitem[\protect\citeauthoryear{Jorba \& Masdemont}{Jorba \&
  Masdemont}{1999}]{Jorba1999}
Jorba A.,  Masdemont J.,  1999, \mn@doi [Physica D: Nonlinear Phenomena]
  {https://doi.org/10.1016/S0167-2789(99)00042-1}, 132, 189

\bibitem[\protect\citeauthoryear{Lei \& Bo}{Lei \& Bo}{2018}]{Lei2018}
Lei H.,  Bo X.,  2018, Astrophysics and Space Science, 363, 70

\bibitem[\protect\citeauthoryear{Maciejewski}{Maciejewski}{1995}]{Maciejewski1995}
Maciejewski A.~J.,  1995, CeMDA, 63, 1

\bibitem[\protect\citeauthoryear{{Margot}, {Pravec}, {Taylor}  \&
  {Carry}}{{Margot} et~al.}{2015}]{Margot2015}
{Margot} J.~L.,  {Pravec} P.,  {Taylor} P.,   {Carry} B.~and{Jacobson} S.,
  2015, {Asteroid Systems: Binaries, Triples, and Pairs}.
pp 355--374, \mn@doi{10.2458/azu_uapress_9780816532131-ch019}

\bibitem[\protect\citeauthoryear{McMahon~J.W.}{McMahon~J.W.}{2013}]{McMahon2013}
McMahon~J.W. S.~D.,  2013, CeMDA, 115, 365

\bibitem[\protect\citeauthoryear{Mcmahon \& Scheeres}{Mcmahon \&
  Scheeres}{2010}]{Mcmahon2010}
Mcmahon J.,  Scheeres D.,  2010, \icarus, 209, 494

\bibitem[\protect\citeauthoryear{Michel et~al.,}{Michel
  et~al.}{2016}]{Michel2016}
Michel P.,  et~al., 2016, \mn@doi [Advances in Space Research]
  {10.1016/j.asr.2016.03.031}, 57, 2529

\bibitem[\protect\citeauthoryear{Murray \& Dermott}{Murray \&
  Dermott}{1999}]{Murray1999}
Murray C.~D.,  Dermott S.~F.,  1999, Solar system dynamics

\bibitem[\protect\citeauthoryear{{Naidu} \& {Margot}}{{Naidu} \&
  {Margot}}{2015}]{Naidu2015}
{Naidu} S.~P.,  {Margot} J.-L.,  2015, \mn@doi [\aj]
  {10.1088/0004-6256/149/2/80}, \href
  {https://ui.adsabs.harvard.edu/abs/2015AJ....149...80N} {149, 80}

\bibitem[\protect\citeauthoryear{{Noyelles}, {Frouard}, {Makarov}  \&
  {Efroimsky}}{{Noyelles} et~al.}{2014}]{Noyelles2014}
{Noyelles} B.,  {Frouard} J.,  {Makarov} V.~V.,   {Efroimsky} M.,  2014,
  \mn@doi [\icarus] {10.1016/j.icarus.2014.05.045}, \href
  {https://ui.adsabs.harvard.edu/abs/2014Icar..241...26N} {241, 26}

\bibitem[\protect\citeauthoryear{Pravec et~al.,}{Pravec
  et~al.}{2016}]{Pravec2016}
Pravec P.,  et~al., 2016, \icarus, 267, 267

\bibitem[\protect\citeauthoryear{Quillen, Nichols-Fleming, Chen  \&
  Noyelles}{Quillen et~al.}{2017}]{Quillen2017}
Quillen A.~C.,  Nichols-Fleming F.,  Chen Y.~Y.,   Noyelles B.,  2017, \icarus,
  293, 94

\bibitem[\protect\citeauthoryear{Scheeres}{Scheeres}{1994}]{Scheeres1994}
Scheeres D.~J.,  1994, \icarus, 110, 225

\bibitem[\protect\citeauthoryear{Scheeres}{Scheeres}{2004}]{Scheeres2004}
Scheeres D.~J.,  2004, \mn@doi [Ann N Y Acad Sci] {10.1196/annals.1311.006},
  1017, 81

\bibitem[\protect\citeauthoryear{Scheeres}{Scheeres}{2009}]{Scheeres2009}
Scheeres D.~J.,  2009, CeMDA, 104, 103

\bibitem[\protect\citeauthoryear{Scheeres et~al.,}{Scheeres
  et~al.}{2006}]{Scheeres2006Dynamical}
Scheeres D.~J.,  et~al., 2006, \mn@doi [\sci] {10.1126/science.1133599}, 314,
  1280

\bibitem[\protect\citeauthoryear{Shi~Y.}{Shi~Y.}{2017}]{Yu2017}
Shi~Y. Wang~Y. X.~S.,  2017, CeMDA, pp 1--14

\bibitem[\protect\citeauthoryear{{Walsh} \& {Jacobson}}{{Walsh} \&
  {Jacobson}}{2015}]{Walsh2015}
{Walsh} K.~J.,  {Jacobson} S.~A.,  2015, {Formation and Evolution of Binary
  Asteroids}.
pp 375--393, \mn@doi{10.2458/azu_uapress_9780816532131-ch020}

\bibitem[\protect\citeauthoryear{Wang \& Xu}{Wang \& Xu}{2018}]{Wang2018}
Wang Y.,  Xu S.,  2018, Astrodynamics, 2, 53

\bibitem[\protect\citeauthoryear{Wisdom, Peale  \& Mignard}{Wisdom
  et~al.}{1984}]{Wisdom1984}
Wisdom J.,  Peale S.~J.,   Mignard F.,  1984, \icarus, 58, 137

\bibitem[\protect\citeauthoryear{Xin, Scheeres  \& Hou}{Xin
  et~al.}{2016}]{Xin2016}
Xin X.,  Scheeres D.~J.,   Hou X.,  2016, CeMDA, 126, 405

\bibitem[\protect\citeauthoryear{Ćuk \& Burns}{Ćuk \& Burns}{2005}]{Cuk2005}
Ćuk M.,  Burns J.~A.,  2005, \icarus, 176, 418

\makeatother
\end{thebibliography}




\appendix

\section{Elements of the matrix \textit{A}}
\label{sec:AppendA}
	\begin{equation}
		\begin{array}l
			a_{31} =\frac{(I_z^{B}\dot{\theta}_0+K)^2}{(m{r}_0^2+I_z^{B})^2}-4\frac{m(I_z^{B}\dot{\theta}_0+K)^2{r}_0^2}{(m{r}_0^2+I_z^{B})^3}+\frac{2}{{r}_0^3}+6\frac{A_1+A_2\cos(2\theta_0)}{{r}_0^5}\\
			a_{32} =\frac{3A_2\sin(2\theta_0)}{{r}_0^4}\\
			a_{33} = 0\\
			a_{34} =\frac{2{r}_0I_z^{B}(I_z^{B}\dot{\theta}_0+K)}{(m{r}_0^2+I_z^{B})^2}\\
			a_{41} =\frac{2\dot{{r}}_0(3m{r}_0^3+I_z^{B})(I_z^{B}\dot{\theta}_0+K)}{(m{r}_0^2+I_z^{B})^2{r}_0^2}+\frac{A_2(3m{r}_0^3+5I_z^{B})\sin(\theta_0)}{I_z^{B}{r}_0^6}\\
			a_{42} =-\frac{2I_z^tA_2\cos(2\theta_0)}{I_z^{B}{r}_0^5}\\
			a_{43} =-\frac{2(I_z^{B}\dot{\theta}_0+K)}{{r}_0(m{r}_0^2+I_z^{B})}\\
			a_{44} =-\frac{2\dot{{r}}_0I_z^{B}}{I_z^t{r}_0}
		\end{array}	\notag
	\label{con:varMC}
	\end{equation}


\bsp	
\label{lastpage}
\end{document}